\documentclass[onecolumn]{gji}

\usepackage[dvips]{graphicx}

\usepackage{natbib}

\usepackage{amsmath}

\begin{document}

\title{Global Gravity Inversion of Bodies with Arbitrary Shape}

\author[Pasquale Tricarico]{Pasquale Tricarico \\ Planetary Science Institute, 1700 E.~Ft.~Lowell Rd., Suite 106, Tucson, AZ 85719, USA}

\maketitle

\begin{summary}
Gravity inversion allows us to constrain the interior mass distribution 
of a planetary body using the observed shape, rotation, and gravity.
Traditionally, techniques developed for gravity inversion can be divided into 
Monte Carlo methods, matrix inversion methods, and spectral methods.
Here we employ both matrix inversion and Monte Carlo in order to 
explore the space of exact solutions,
in a method which is particularly suited for arbitrary shape bodies.
We expand the mass density function using orthogonal polynomials,
and map the contribution of each term to the global gravitational field generated.
This map is linear in the density terms, and can be pseudo-inverted
in the under-determined regime using 
QR decomposition, to obtain a basis of the affine space of exact interior structure solutions.
As the interior structure solutions are degenerate, 
assumptions have to be made in order to control their properties,
and these assumptions can be transformed into scalar functions and
used to explore the solutions space using Monte Carlo techniques.
Sample applications show that the range of solutions tend to converge towards the nominal one
as long as the generic assumptions made are correct,
even in the presence of moderate noise.
We present the underlying mathematical formalism and an analysis of how to impose specific features
on the global solution, including uniform solutions, gradients, and layered models.
Analytical formulas for the computation of the relevant quantities when the shape is represented 
using several common methods are included in the Appendix.
\end{summary}

\section{Introduction}

The range of geophysical features expressed by planetary bodies
is the result of their formation mechanism, thermal evolution,
collisional, dynamical and rotational history, as well as surface processes.
Their interiors encompass a wide range of structures,
loosely correlated to the body size.
Small bodies are strength dominated and characterized by irregular shapes, high topographic reliefs, 
terrain slopes close to the angle of repose of their surface materials,
and mass density distribution close to homogeneous, in many cases associated with high porosity \citep{1973Icar...18..612J,2002aste.conf..463A,2003P&SS...51..443B}.
Large moons and terrestrial planets on the other hand are gravity dominated,
tend to have spheroidal shapes, low topographic relief, shallow terrain slopes,
and their interior structure is differentiated into several layers 
\citep{1973Icar...18..612J,Wieczorek_2007}.

Planetary bodies with intermediate sizes may be expressing some of the most intriguing characteristics,
and have been the targets of recent robotic missions. 
Their size is such that strength and gravity cause comparable effects,
their thermal evolution might have lead to complete or partial melting early after their formation
\citep{1955PNAS...41..127U,1995Metic..30..365M,1998Icar..134..187G},
their collisional history may include the formation of impact craters with sizes comparable in scale with the body size.
Their interior structure is affected by all these factors.
Recent robotic exploration 
is shedding some light on their interior structure.
In 2010 the Rosetta spacecraft flew by 21~Lutetia, a main belt asteroid with approximate dimensions $121 \times 101 \times 75$~km \citep{2011Sci...334..491P,2011Sci...334..487S}.
The estimated bulk density of $3.4 \pm 0.3$~g/cm$^3$
exceeds that of most known chondritic meteorite groups,
and may indicate that 21~Lutetia is partially differentiated \citep{2012P&SS...66..137W}.
In 2011 the Dawn spacecraft started a year long exploration of 4~Vesta,
also a main belt asteroid, with an oblate spheroid shape with approximate dimensions $570 \times 570 \times 458$~km \citep{2012Sci...336..687J},
and initial gravity results are consistent with the presence of 
a 110~km radius iron core \citep{2012Sci...336..684R}.
As we move forward in the exploration of this intermediate class of planetary bodies,
and we attempt to model their interior structure, 
we need to develop an approach which is capable of generating the entire range of 
solutions which are compatible with the observed data, and which
satisfy a set of assumptions. The assumptions have to be chosen carefully,
and the effect of each one should be clearly accounted for in the analysis of the results.

If we restrict ourselves to gravity inversion methods which work directly in the mass density space,
then the mass distribution can be modeled with a given number of free parameters,
and solutions can be found by sampling the parameters space 
while comparing the observed gravity with the gravity field generated by the model.
This is also commonly referred to as forward modeling.
In \cite{1977BGeod..51..227D} and \cite{1993MGeo...18...99B} the mass distribution of a sphere and the gravity field generated 
are both modeled using orthogonal functions,
and a direct relation is determined between the coefficients of the expansion of mass and gravity.
This is then extended in \cite{2005GeoJI.162...32C} using a more complex kernel function.
Polyhedral shapes with homogeneous mass density are used to model the gravity of a body 
\citep{1976Geop...41.1353B,1997CeMDA..65..313W,2000P&SS...48..965S},
or to assign a shape to a layer or other interior component of the body 
\citep{1989JGR....94.7555R,2007Icar..192..150H},
and then changes to either the shape or the density of some of the elements allow the modification
of the generated gravity and its comparison to the observed gravity.
In \cite{1998Geop...63..109L} the interior is modeled using a large number of small brick-like elements,
with a density assigned to each element.
Other methods include 
free-positioned point mass modeling \citep{1978JGR....83..841R,1983MZPE.1214.....B,1991dgpf.conf..484B,1993BGeod..67...31L},
radial multipole methods \citep{2008StGG...52..287T,2008JGeod..82..457K}, 
wavelet methods \citep{2005GeoJI.163..875C,2008InvPr..24d5019M}, 
and spline methods \citep{2008GeoJI.173....1M,Berkel_Michel_2010}.
Common to all these techniques is the necessity to effectively sample the parameters space
and generate models which agree with the observations,
and the Monte Carlo approach provides several methods to achieve this goal
in a wide range of scenarios \citep{1995JGR...10012431M,2002RvGeo..40.1009S}.

Forward modeling using Monte Carlo sampling can be very time consuming
when the number of free parameters in the model is large. 
This prompts us to find a transformation of the parameters in the model,
which allows us to work in a space of exact models, 
i.e.~models which generate exactly the gravity field observed
up to a given degree, which defines the dimensionality of the problem.
This method involves matrix inversion in an under-constrained interior model,
and as we show in the remainder of this manuscript,
it allows us to efficiently explore the space of exact solutions,
while accounting for an arbitrary number of assumptions made on the properties of the solutions.
After introducing the common notation for gravity and the relation between 
generalized moments of inertia (GMoI) and gravitational field expansion in \S\ref{sec:modeling},
we present the details of this technique in \S\ref{sec:MatrixTheory}, and then apply it to a sample body 
in three different mass distribution cases in \S\ref{sec:sample}, including one with noise.
A discussion including the role of uncertainty in the observed quantities is provided in \S\ref{sec:discussion}.

\section{Modeling the Gravitational Field}
\label{sec:modeling}

The gravitational potential of a planetary body
with arbitrary shape and mass distribution
can be described using the spherical harmonics series \citep{1966tsga.book.....K,1995geph.conf....1Y}:
\begin{equation}
U(r,\theta,\phi) = \frac{G M}{r} \sum_{l=0}^{\infty} \sum_{m=0}^{l} \left(\frac{r_0}{r}\right)^{l} P_{lm}(\cos\theta) \left( C_{lm} \cos m\phi + S_{lm} \sin m\phi \right) \label{eq:potential}
\end{equation}
where $G$ is the universal gravitational constant,
$M$ is the total mass of the body,
$\{r,\theta,\phi\}$ are the body-fixed barycentric spherical coordinates 
(radius, co-latitude, longitude)
of the point where the potential $U$ is computed,
$r_0$ is an arbitrary reference radius typically close to the equatorial radius,
$P_{lm}(\cos\theta)$ is the associate Legendre function.
Using this series, the Stokes coefficients $\{C_{lm},S_{lm}\}$ uniquely characterize the potential of the body.
In cases where the mass density distribution $\varrho(r,\theta,\phi)$ of a body is known,
the coefficients $\{C_{lm},S_{lm}\}$ 
can be determined by integrating 
over the volume $V$ of the body \citep{1995geph.conf....1Y}:
\begin{align}
C_{lm} &= \frac{(2-\delta_{m,0})}{M} \frac{(l-m)!}{(l+m)!} \int_V \varrho(r,\theta,\phi) \left( \frac{r}{r_0} \right)^{l} P_{lm}(\cos\theta) \cos m\phi \ \ \text{d}V \label{eq:Clm} \\
S_{lm} &= \frac{(2-\delta_{m,0})}{M} \frac{(l-m)!}{(l+m)!} \int_V \varrho(r,\theta,\phi) \left( \frac{r}{r_0} \right)^{l} P_{lm}(\cos\theta) \sin m\phi \ \ \text{d}V \label{eq:Slm}
\end{align}
The Stokes coefficients can be more readily compared and used in numerical
work when their magnitude is normalized \citep{1995geph.conf....1Y}:
\begin{align}
\{\bar{C}_{lm},\bar{S}_{lm}\} & = N_{lm} \{C_{lm},S_{lm}\} \\
\label{eq:norm} N_{lm} & = \sqrt{\frac{(l+m)!}{(2-\delta_{m,0})(2l+1)(l-m)!}}
\end{align}
where $N_{lm}$ is the normalization factor.

\begin{figure}
\begin{center}
\includegraphics*[bb=100 420 515 690,width=\columnwidth]{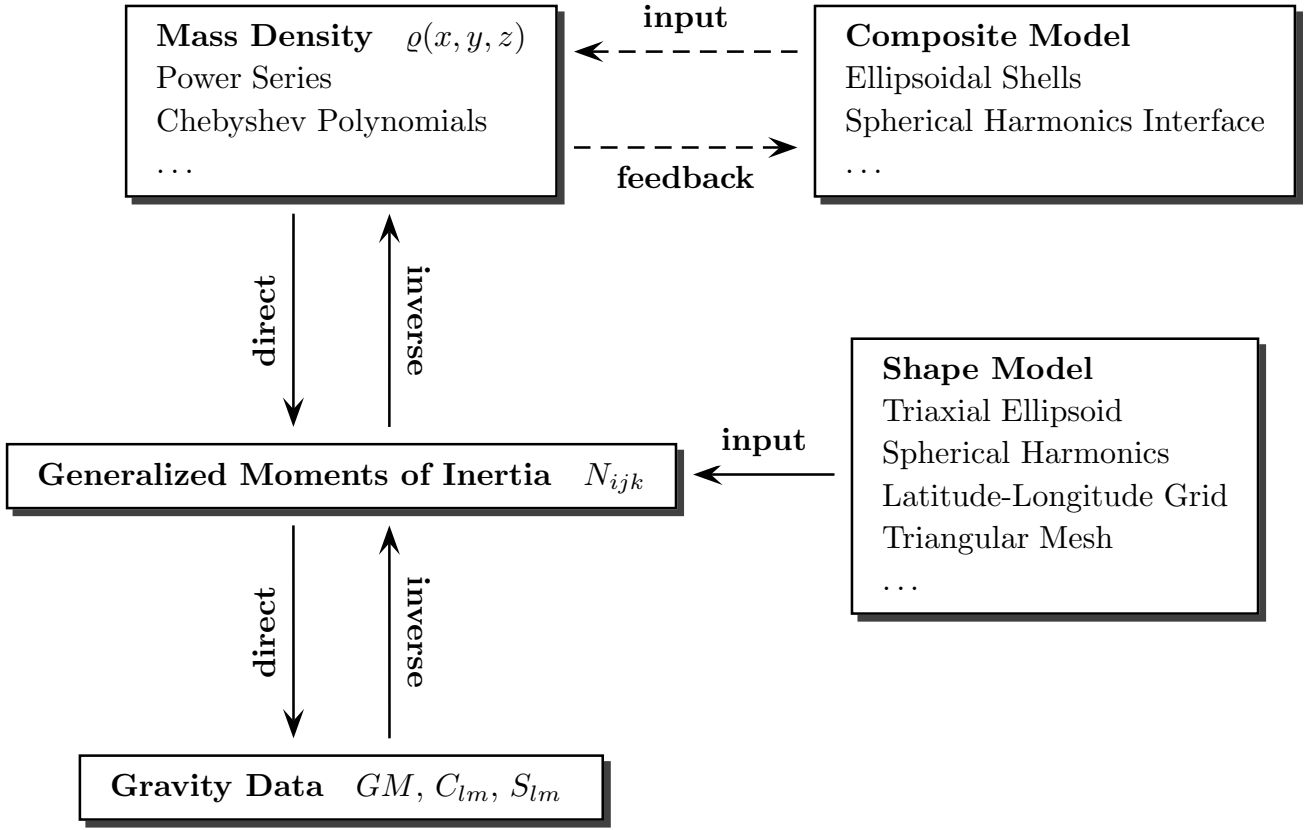}
\end{center}
\caption{
The gravity data (total mass, spherical harmonics expansion of the gravity potential) depends linearly on the
generalized moments of inertia, which combine spatial data (shape model) and mass density data.
The direct and inverse problems are marked by the corresponding arrows,
and the shape model is considered a static input in the computation of the generalized moments of inertia.
Composite models such as layered mass density distributions (\S\ref{sec:composite})
can be used to describe complex interior structure models,
and their characteristics are typically fixed as in a static input, 
but can also be iteratively updated via a feedback from the main mass density.
}
\label{fig:diagram}
\end{figure}

It is useful to introduce the generalized moments of inertia (GMoI),
which have the form \citep{1988CeMec..44...49P}:
\begin{equation}
{\cal M}_{ijk} = \int_{V} x^i y^j z^k \varrho(x,y,z) \ \mbox{d}V
\label{eq:Mijk}
\end{equation}
and can be used in an adimensional form
${\cal N}_{ijk}$ 
\citep{2008CeMDA.100..319T}:
\begin{equation}
\displaystyle
{\cal N}_{ijk} \equiv
\frac{\displaystyle \int_{V} \frac{x^i y^j z^k}{r_0^{i+j+k}} \varrho(x,y,z) \ {\text d}V}{\displaystyle \int_{V} \varrho(x,y,z) \ {\text d}V} =
\frac{1}{r_0^{i+j+k}} \frac{{\cal M}_{ijk}}{{\cal M}_{000}}
\label{eq:Nijk}
\end{equation}
where ${\cal M}_{000} \equiv M$ is the total mass of the body.
The GMoI have the appealing characteristic of
combining under a single variable the inertial and spatial properties of a body. 
The degree of a GMoI goes from zero to infinity, capturing all scales and wavelengths.
The GMoI are defined in cartesian coordinates, usually in the body-fixed reference frame,
and the dependency on the inertial component
can be kept implicit by using a polynomial expansion of the mass density distribution.
This brings one more advantage, that is that the GMoI capture both inertial and spatial information,
but depend explicitly only on the spatial information,
and as such, can be used as an intermediate step 
when solving the direct or inverse problem, see Figure~\ref{fig:diagram}.
The direct problem has already been solved in \cite{2008CeMDA.100..319T}, 
where the Stokes coefficients are expressed in terms of the GMoI,
and including the normalization factors \citep{1995geph.conf....1Y} we have:
\begin{align}
\label{eq:CS}
\begin{split}
\bar{C}_{lm} = {} &
\frac{1}{2^l}
\sqrt{\frac{(l-m)!}{(l+m)!}\frac{(2-\delta_{0m})}{(2l+1)}} 
\sum_{p=0}^{\lfloor l/2 \rfloor} 
\sum_{q=0}^{\lfloor m/2 \rfloor} 
(-1)^{p+q}
\binom{l}{p} 
\binom{2l-2p}{l} 
\binom{m}{2q} 
\\
&
(l - m - 2 p + 1)_{m}
\sum_{\nu_x=0}^p
\sum_{\nu_y=0}^{p-\nu_x}
\frac{p!}{\nu_x!\nu_y!(p-\nu_x-\nu_y)!}
{\cal N}_{m-2q+2\nu_x,2q+2\nu_y,l-m-2\nu_x-2\nu_y} 
\end{split} \\
\begin{split}
\bar{S}_{lm} = {} &
\frac{1}{2^l}
\sqrt{\frac{(l-m)!}{(l+m)!}\frac{(2-\delta_{0m})}{(2l+1)}} 
\sum_{p=0}^{\lfloor l/2 \rfloor} 
\sum_{q=0}^{\lfloor (m-1)/2 \rfloor}   
(-1)^{p+q}
\binom{l}{p} 
\binom{2l-2p}{l} 
\binom{m}{2q+1} 
\\
&
(l - m - 2 p + 1)_{m} 
\sum_{\nu_x=0}^p
\sum_{\nu_y=0}^{p-\nu_x}
\frac{p!}{\nu_x!\nu_y!(p-\nu_x-\nu_y)!}
{\cal N}_{m-2q-1+2\nu_x,2q+1+2\nu_y,l-m-2\nu_x-2\nu_y}
\end{split}
\end{align}
and the terms up to degree 4 are provided in Table~\ref{tab:conv}.

The mass density of the body $\varrho(x,y,z)$ can be modeled using a generic orthogonal polynomial basis 
$P_{w_1 w_2 \cdots}(x,y,z)$ or alternatively using a power series, to obtain the expansions:
\begin{align}
\begin{split}
\varrho(x,y,z) 
& = \sum_{w_1 w_2 \cdots=0}^{\{w_1 w_2 \cdots\}_{\max}} c'_{w_1 w_2 \cdots} P_{w_1 w_2 \cdots}(x,y,z) \\
& = \sum_{i_\varrho j_\varrho k_\varrho=0}^{i_\varrho+j_\varrho+k_\varrho=N_\varrho} c_{i_\varrho j_\varrho k_\varrho} \dfrac{x^{i_\varrho} y^{j_\varrho} z^{k_\varrho}}{r_0^{i_\varrho+j_\varrho+k_\varrho}}
\end{split}
\label{eq:generic_density_expansion}
\end{align}
where $w_1 w_2 \cdots$ is a set of variables which are characteristic of the chosen orthogonal polynomials, 
with $c'_{w_1 w_2 \cdots}$ the corresponding coefficients,
which become $c_{i_\varrho j_\varrho k_\varrho}$ in the specific case of a power series expansion.
The maximum degree of the power series expansion of the mass density distribution is $N_\varrho$.
The dependence on $(x,y,z)$ can be implicit in $P_{w_1 w_2 \cdots}(x,y,z)$, 
but needs to be made explicit in order to easily relate the density coefficients to the GMoI expansion.
In this manuscript we assume that this is always possible, 
see \S\ref{sec:MDrepMAIN} for an example.
Additionally, we note that since this expansion does not include explicitly density discontinuities
such as in a differentiated body,
these can be approximated using high degree expansions, or can be included exactly 
and described in \S\ref{sec:composite}.
The GMoI expansion Eq.~\eqref{eq:Nijk} using Eq.~\eqref{eq:generic_density_expansion} becomes:
\begin{align}
\label{eq:Nijk_expand_generic} 
\begin{split}
{\cal N}_{ijk} 
& = \dfrac{r_0^3}{M} \sum_{i_\varrho j_\varrho k_\varrho=0}^{i_\varrho+j_\varrho+k_\varrho=N_\varrho} c_{i_\varrho j_\varrho k_\varrho} \int_{V} \dfrac{x^{i+i_\varrho} y^{j+j_\varrho} z^{k+k_\varrho}}{r_0^{i+i_\varrho+j+j_\varrho+k+k_\varrho}} \ \dfrac{{\text d}V}{r_0^3} \\
& = \dfrac{r_0^3}{M} \sum_{i_\varrho j_\varrho k_\varrho=0}^{i_\varrho+j_\varrho+k_\varrho=N_\varrho} c_{i_\varrho j_\varrho k_\varrho} \Phi_{i+i_\varrho,j+j_\varrho,k+k_\varrho}(r_0,V) \\
\end{split}
\end{align}
where the integral 
$\Phi$
depends only on the shape of the body.
In general we refer to these volume integrals with:
\begin{align}
\label{eq:integral_def} 
\Phi_{ijk}(r_0,V) = \int_{V} \dfrac{x^{i} y^{j} z^{k}}{r_0^{i+j+k}} \ \dfrac{{\text d}V}{r_0^3}
\end{align}
or more shortly to $\Phi_{ijk}$ when there is no ambiguity as to what $r_0$ and $V$ are,
and in \S\ref{sec:SMrepMAIN} we compute explicitly $\Phi_{ijk}$
for several common shape model representations.
Note how the GMoI expression in Eq.~\eqref{eq:Nijk_expand_generic} mixes contributions from different $c_{i_\varrho j_\varrho k_\varrho}$.
Finally, by plugging Eq.~\eqref{eq:Nijk_expand_generic} into Eq.~\eqref{eq:CS} we can 
map the mass density coefficients $c_{i_\varrho j_\varrho k_\varrho}$
to the normalized spherical harmonics coefficients of the gravitational potential.

\section{Linear Inversion}
\label{sec:MatrixTheory}

The linear relation between mass density expansion coefficients and gravitational field spherical harmonics
can be described using matrices.
Here we describe the passage to matrix notation,
and then show how in order to obtain exact solutions which explore the entire solutions space, 
we have to work with an under-determined problem.
The basis for the null space of degenerate exact solutions
is generated using QR decomposition \citep{1996maco.book.....G} of the main matrix connecting density and gravity, 
and a reference solution is obtained using a pseudo-inverse matrix.

For the following part it is important to deal with 2D matrices,
and in order to do so, we use a unique index for $lm$ of gravity 
and one for $i_\varrho j_\varrho k_\varrho$ of density.
We have then a total of $\dim(lm) = (l_{\max}+1)^2$ terms of which 
$(l_{\max}+1)(l_{\max}+2)/2$ are $\bar{C}_{lm}$ and 
$l_{\max}(l_{\max}+1)/2$ are $\bar{S}_{lm}$.
For $ijk$ we have $\dim(ijk) = (l_{\max}+1)(l_{\max}+2)(l_{\max}+3)/6$, where $l_{\max} = (i+j+k)_{\max}$.
The case of $i_\varrho j_\varrho k_\varrho$ is similar to $ijk$, with $N_\varrho$ replacing $l_{\max}$. 
Note how the two-index size grows quadratically while the three-index size grows cubically,
which has the consequence that if gravity and GMoI have the same maximum degree,
as is advisable in order to have comparable spatial resolution,
the GMoI has more degrees of freedom than gravity,
causing the inverse problem to be under-determined.

We can now rewrite Eq.~\eqref{eq:CS} using the more compact notation:
\begin{equation}
\label{eq:compact_CS}
\begin{bmatrix}
\bar{C}_{lm} \\ 
\bar{S}_{lm} \\
\end{bmatrix}
= 
{\mathbf{A}}({{\cal N}\rightarrow{\bar{C}\bar{S}}}) 
\begin{bmatrix} {\cal N}_{ijk} \end{bmatrix}
\end{equation}
where
$\begin{bmatrix}
\bar{C}_{lm} \\ 
\bar{S}_{lm} \\
\end{bmatrix}$
and
$\begin{bmatrix} {\cal N}_{ijk} \end{bmatrix}$
are column vectors with elements:
\begin{align}
\label{eq:explicit_CS_vector}
\begin{split}
\begin{bmatrix}
\bar{C}_{lm} \\ 
\bar{S}_{lm} \\
\end{bmatrix}
& = 
\begin{bmatrix}
\bar{C}_{00} \\ 
\bar{C}_{10} \\ 
\bar{C}_{11} \\
\bar{S}_{11} \\
\bar{C}_{20} \\
\bar{C}_{21} \\
\bar{S}_{21} \\
\dots \\ 
\bar{C}_{l_{\text{max}}l_{\text{max}}} \\
\bar{S}_{l_{\text{max}}l_{\text{max}}} \\
\end{bmatrix}
\ \ \text{and} \ \ 
\begin{bmatrix} {\cal N}_{ijk} \end{bmatrix} 
= 
\begin{bmatrix} 
{\cal N}_{000} \\
{\cal N}_{001} \\
{\cal N}_{010} \\ 
\dots \\
{\cal N}_{l_{\max}00} \\
\end{bmatrix}
\end{split}
\end{align}
Here we use the notation ${\mathbf{A}}(x \rightarrow y)$ to indicate that the matrix
${\mathbf{A}}$ transforms the vector $\begin{bmatrix} x \end{bmatrix}$ into 
$\begin{bmatrix} y \end{bmatrix} = {\mathbf{A}}(x \rightarrow y) \begin{bmatrix} x \end{bmatrix}$. 
Note that
$\bar{C}_{00} = 1$,
$\bar{C}_{10} = 0$,
$\bar{C}_{11} = 0$,
$\bar{S}_{11} = 0$
in the center of mass reference frame,
so their inclusion could appear as redundant in this context
as they do not convey any specific information about the gravity field studied.
But it turns out that 
their inclusion is necessary in order to guarantee that the total mass is conserved in every solution (degree 0),
and that the position of the center of mass is conserved in every solution (degree 1),
which in general may be offset from the origin of a body-fixed reference frame.
The matrix ${\mathbf{A}}({{\cal N}\rightarrow{\bar{C}\bar{S}}})$
with coefficients from Eq.~\eqref{eq:CS}
transforms the $\begin{bmatrix} {\cal N}_{ijk} \end{bmatrix}$ vector into the 
$\begin{bmatrix}
\bar{C}_{lm} \\ 
\bar{S}_{lm} \\
\end{bmatrix}$ 
vector.
Similarly, we can introduce a matrix ${\mathbf{A}}(c\rightarrow{\cal N})$ with coefficients from Eq.~\eqref{eq:Nijk_expand_generic} to obtain:
\begin{align}
\label{eq:compact_CS_with_density_coeff}
\begin{split}
\begin{bmatrix}
\bar{C}_{lm} \\ 
\bar{S}_{lm} \\
\end{bmatrix} 
& = 
{\mathbf{A}}({{\cal N}\rightarrow{\bar{C}\bar{S}}}) 
{\mathbf{A}}(c\rightarrow{\cal N}) 
\begin{bmatrix} c_{i_\varrho j_\varrho k_\varrho} \end{bmatrix} \\
& = 
{\mathbf{A}}(c\rightarrow{\bar{C}\bar{S}}) 
\begin{bmatrix} c_{i_\varrho j_\varrho k_\varrho} \end{bmatrix}
\end{split}
\end{align}
with the vector $\begin{bmatrix} c_{i_\varrho j_\varrho k_\varrho} \end{bmatrix}$ containing all the density coefficients, 
and where the matrix product 
${\mathbf{A}}(c\rightarrow{\bar{C}\bar{S}}) = {\mathbf{A}}({{\cal N}\rightarrow{\bar{C}\bar{S}}}) {\mathbf{A}}(c\rightarrow{\cal N})$
transforms the density distribution into the gravitational potential.
The matrix ${\mathbf{A}}(c\rightarrow{\bar{C}\bar{S}})$ has dimensions $(l_{\max}+1)^2 \times (N_\varrho+1)(N_\varrho+2)(N_\varrho+3)/6$.
The direct problem of determining the gravity and inertial properties of a body with given shape and 
mass distributions consists in simply plugging the $\begin{bmatrix} c_{i_\varrho j_\varrho k_\varrho} \end{bmatrix}$ coefficients into Eq.~\eqref{eq:compact_CS_with_density_coeff}.

Solving the inverse problem is equivalent to inverting the matrix ${\mathbf{A}}(c\rightarrow{\bar{C}\bar{S}})$
to obtain a matrix ${\mathbf{A}}({\bar{C}\bar{S}}\rightarrow c)$, but in general this is not possible.
Matrix inversion is well defined only for a square matrix which is not singular,
and that would provide us a single solution 
\begin{align}
\begin{bmatrix} c_{i_\varrho j_\varrho k_\varrho} \end{bmatrix} _\text{ref.}
= 
{\mathbf{A}}({\bar{C}\bar{S}}\rightarrow c) 
\begin{bmatrix}
\bar{C}_{lm} \\ 
\bar{S}_{lm} \\
\end{bmatrix}
\end{align}
which would tell us very little about the family of solutions which are compatible with the observed shape and gravity data.
Our alternatives are to work with an over-constrained problem 
($\dim(i_\varrho j_\varrho k_\varrho) < (l_{\max}+1)^2$)
or an under-constrained problem 
($\dim(i_\varrho j_\varrho k_\varrho) > (l_{\max}+1)^2$).
In the over-constrained gravity inversion problem, the number of degrees of freedom
in the mass density expansion is smaller than the constrains from the gravity potential data.
This does not generate exact solutions, but least square solutions, and we are not going to cover that here,
please see \cite{Snieder_Trampert_1999} for a discussion.

In the under-constrained gravity inversion problem, we have infinite solutions, if any,
all with the property of generating exactly the observed gravitational field used as input.
This allows us to explore the space of exact solutions while searching 
for a sub-set of solutions which satisfy a given set of assumptions,
and in order to proceed,
we need to determine an initial reference solution 
$\begin{bmatrix} c_{i_\varrho j_\varrho k_\varrho} \end{bmatrix} _\text{ref.}$, 
and find a set of vectors which form an orthonormal basis of the null space.
The reference solution 
$\begin{bmatrix} c_{i_\varrho j_\varrho k_\varrho} \end{bmatrix}_\text{ref.}$ 
can be obtained using the pseudo-inverse matrix.
If we have $b_m = {\mathbf A}_{m \times n} x_n$,
the pseudo-inverse matrix
\begin{align}
{\mathbf A}^\dagger_{n \times m} = {\mathbf A}^T_{n \times m} ({\mathbf A} {\mathbf A}^T)_{m \times m}^{-1}
\label{eq:pseudoinverse}
\end{align}
is such that
${\mathbf A}^\dagger_{n \times m} b_m = \tilde x_n$
is a solution, in general different from the initial $x_n$ because of the non-trivial null space.

A real valued rectangular matrix ${\mathbf A}_{m \times n}$ admits a QR decomposition
${\mathbf A}_{m \times n} = {\mathbf Q}_{m \times m}{\mathbf R}_{m \times n}$ 
where ${\mathbf Q}$ is square and orthogonal and ${\mathbf R}$ is right triangular \citep{1996maco.book.....G}.
If ${\mathbf A}_{m \times n}$ with $m<n$ has full row rank=$m$, then 
the null space of ${\mathbf A}$ is non-trivial.
By decomposing ${\mathbf A}^T_{n \times m} = {\mathbf Q}_{n \times n}{\mathbf R}_{n \times m}$ we can obtain from the
${\mathbf Q}$ matrix $m$ columns which are vectors of the orthonormal basis of the range of ${\mathbf A}$, 
and $n-m$ columns which are vectors of the orthonormal basis for the kernel of ${\mathbf A}$ \citep{1996maco.book.....G}.
If we denote with $\begin{bmatrix} u_q \end{bmatrix}$ the kernel basis vectors, 
and with $\begin{bmatrix} c_{i_\varrho j_\varrho k_\varrho} \end{bmatrix} _\text{ref.}$ a reference solution, 
then any other solution has the form 
\begin{equation}
\begin{bmatrix} c_{i_\varrho j_\varrho k_\varrho} \end{bmatrix} 
=
\begin{bmatrix} c_{i_\varrho j_\varrho k_\varrho} \end{bmatrix}_\text{ref.} 
+ 
\sum_q s_q \begin{bmatrix} u_q \end{bmatrix}
\label{eq:gensolref}
\end{equation}
where the sum is over $n-m$ terms,
and $s_q$ are arbitrary factors.
By using Eq.~\eqref{eq:gensolref} we can explore the exact solutions space,
and Monte Carlo techniques provide some of the most efficient methods \citep{1983Sci...220..671K,1995JGR...10012431M,2002RvGeo..40.1009S}.

\section{Application}
\label{sec:sample}

In order to demonstrate this approach,
we study the interior structure of a sample body,
with an arbitrary exterior shape modeled using the 
spherical harmonics coefficients in Table~\ref{tab:SH_body_1},
which correspond to a shape of total dimensions $158.9 \times 113.5 \times 87.1$~km.
This object is assigned a total mass
$M = 1.988692 \times 10^{18}$~kg, and
is studied in three different configurations:
first assuming that the mass is uniformly distributed (\S\ref{sec:uniform});
then assuming that it is composed of three layers (\S\ref{sec:composite});
finally the same layered model but in presence of noise (\S\ref{sec:realistic})
In each case, we first compute gravity coefficients using the assumed mass distribution, 
and then we use the gravity coefficients to try to model the interior structure,
using the approach outlined in \S\ref{sec:MatrixTheory} and ignoring what we already know about it.
The volume integrals as defined in Eq.~\eqref{eq:integral_def} 
are computed using the formalism in \S\ref{sec:SMrepSH},
and the results are listed in Table~\ref{tab:shape_integral_body_1}
and apply to the two noise-free cases.
The GMoI coefficients are then computed using Eq.~\eqref{eq:Nijk_expand_generic},
to obtain the spherical harmonics expansion of the gravitational potential in Table~\ref{tab:gravity_body_1}.
This completes the direct problem (see Figure~\ref{fig:diagram}),
and in the next sections we describe how the inverse problem is solved in each case.

\subsection{Uniform Case}
\label{sec:uniform}

\begin{figure}
\begin{center}
\includegraphics*[bb=310 30 550 435,angle=270,width=0.49\textwidth]{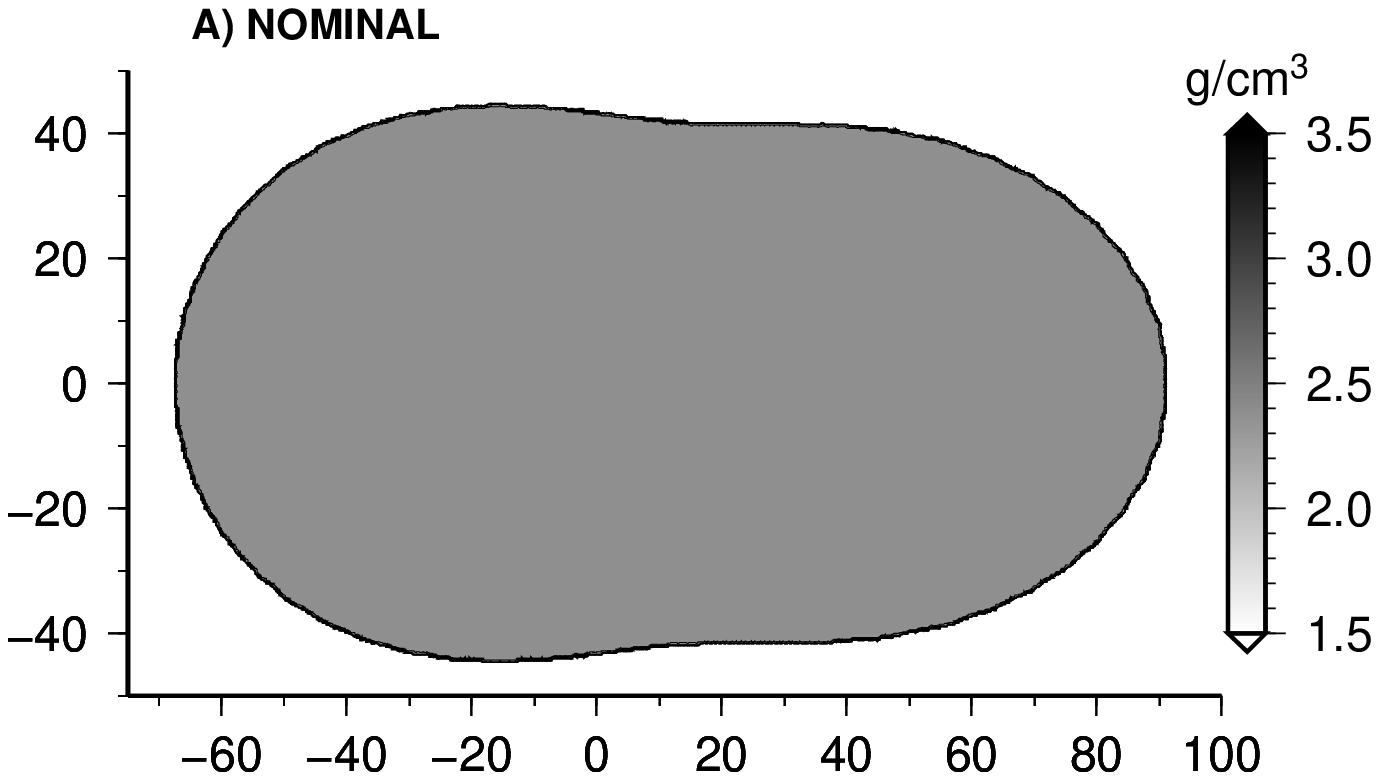}
\includegraphics*[bb=310 30 550 435,angle=270,width=0.49\textwidth]{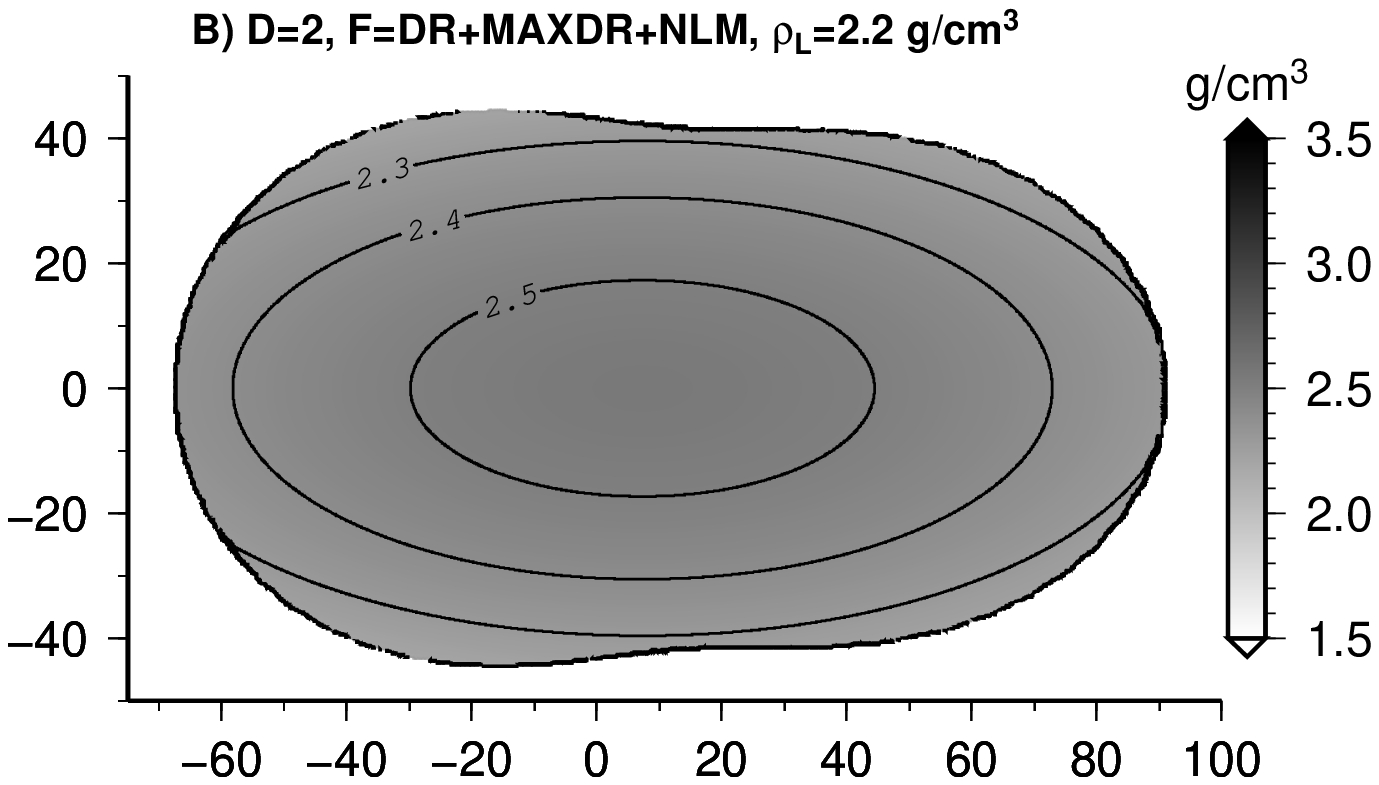}
\includegraphics*[bb=310 30 550 435,angle=270,width=0.49\textwidth]{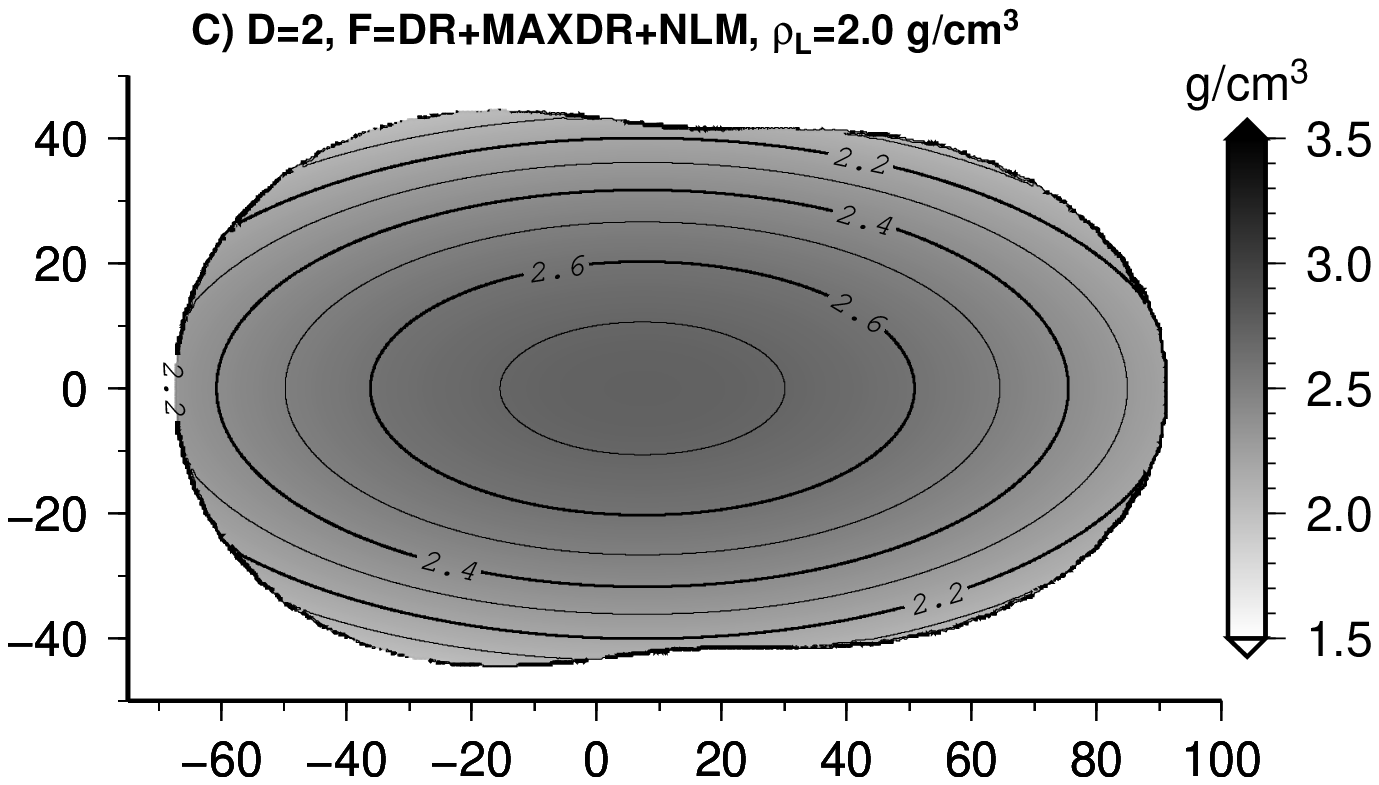}
\includegraphics*[bb=310 30 550 435,angle=270,width=0.49\textwidth]{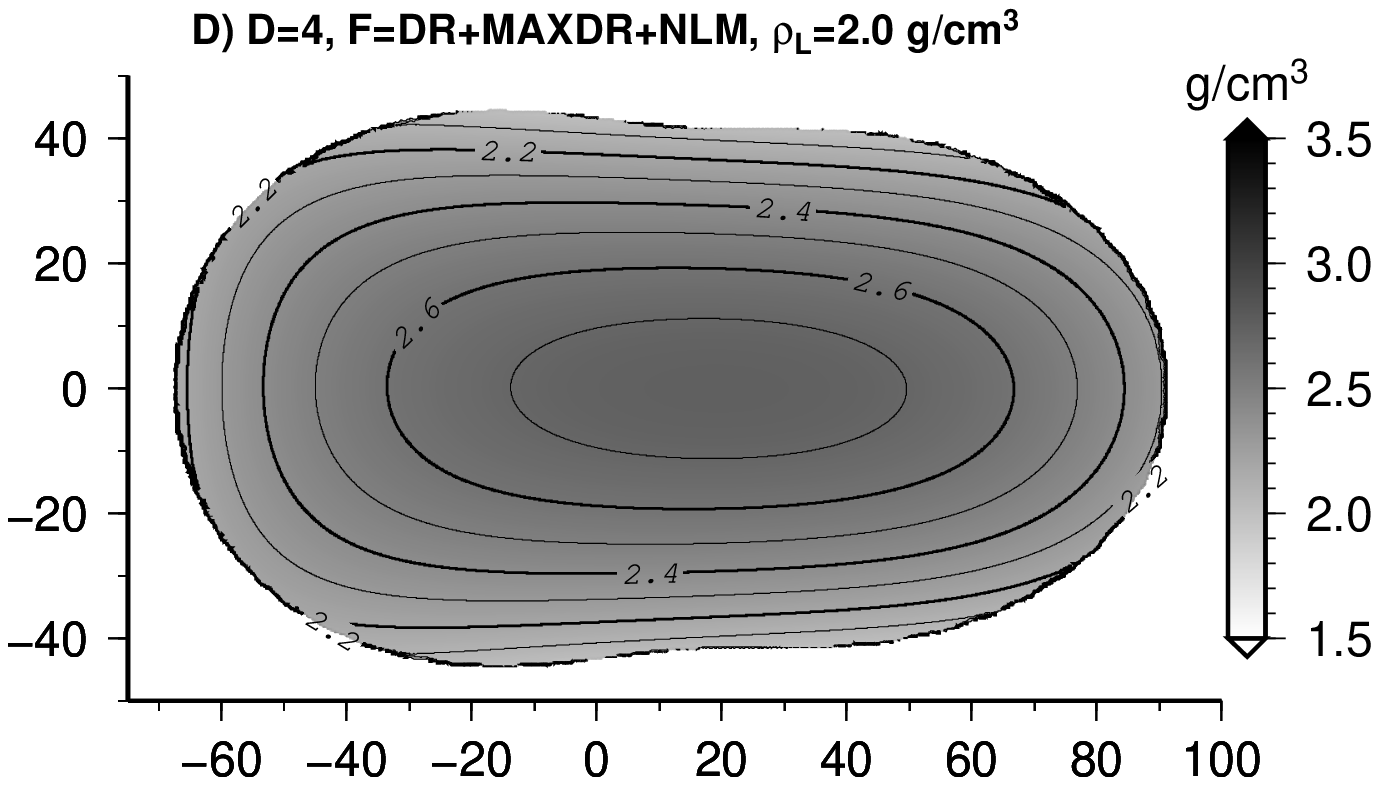}
\end{center}
\caption{Longitudinal section of the sample body in the uniform mass distribution case,
displaying the mass density distribution for the nominal solution (A),
and for other solutions where we varied the number D of degrees of gravity and mass distribution,
the target function F, and the minimum body density $\varrho_L$.
The section is in the $x$ and $z$ axes at $y=0$ of the body-fixed frame, with units in km,
and the density contour lines are at increments of 0.1~g/cm$^3$.
}
\label{fig:sol_U}
\end{figure}

In this first case, the body is assumed to have a uniform mass distribution
equal to the bulk density value,
see Figure~\ref{fig:sol_U}-A.
If we work in the Chebyshev representation of the mass density as described in \S\ref{sec:MDrepMAIN},
instead of the plain power series,
the relation of Eq.~\eqref{eq:Nijk_expand_generic} becomes:
\begin{align}
\begin{bmatrix} {\cal N}_{ijk} \end{bmatrix} 
=
{\mathbf{A}}(c\rightarrow{\cal N})
\begin{bmatrix} c_{i_\varrho j_\varrho k_\varrho} \end{bmatrix}
=
{\mathbf{A}}(c'\rightarrow{\cal N}) 
\begin{bmatrix} c'_{i_\varrho j_\varrho k_\varrho} \end{bmatrix}
\end{align}
where now the $c'$ coefficients are in the Chebyshev representation.
As the body is assumed to have a uniform mass density of $\varrho_{\text{bulk}} = M / (r_0^3 \Phi_{000}) = 2.377647$~g/cm$^3$, 
the only coefficient of the density expansion which is non zero is $c'_{000}=\varrho_{\text{bulk}}$
and the total mass of the body is $M = c'_{000} r_0^3 \Phi_{000} = 1.988692 \times 10^{18}$~kg, 
with $\Phi_{000}$ from Table~\ref{tab:shape_integral_body_1}.
By using Eq.~\eqref{eq:CS} we can obtain the normalized
coefficients $\bar{C}_{lm}$ and $\bar{S}_{lm}$,
and note that $\bar{C}_{11} = 0.047548$, 
which indicates that the origin of the reference system used is not at the center of mass of the body,
but is instead offset by $\Delta_x = \sqrt{3} r_0 \bar{C}_{11} = 8.235548$~km (see Table~\ref{tab:conv}).
We can compute the GMoI in the translated (barycentric) reference frame 
using Eq.~\eqref{eq:Nijk_translation},
and the barycentric coefficients for degree 2 ($N_\varrho=l_{\max}=2$) are:
{
\begin{align}
\begin{bmatrix}
\bar{C}_{00}\\
\bar{C}_{10}\\
\bar{C}_{11}\\
\bar{S}_{11}\\
\bar{C}_{20}\\
\bar{C}_{21}\\
\bar{S}_{21}\\
\bar{C}_{22}\\
\bar{S}_{22}\\
\end{bmatrix}
= 
{\mathbf{A}}(c'\rightarrow{\bar{C}\bar{S}})
\begin{bmatrix}
\varrho_{\text{bulk}}\\
0\\
0\\
0\\
0\\
0\\
0\\
0\\
0\\
0\\
\end{bmatrix}
=
\left[
\begin{array}{r}
 1.000000 \\
 0.000000 \\
 0.000000 \\
 0.000000 \\
-0.022531 \\
 0.000000 \\
 0.000000 \\
 0.027357 \\
 0.000000 \\
\end{array}
\right]
\end{align}
}

The pseudo-inverse of ${\mathbf{A}}(c'\rightarrow{\bar{C}\bar{S}})$ 
can be computed from Eq.~\eqref{eq:pseudoinverse}, 
and by performing QR decomposition (\S\ref{sec:MatrixTheory}) of the transpose matrix ${\mathbf{A}^T}(c'\rightarrow{\bar{C}\bar{S}})$, 
we can obtain the basis vectors of the null space of the matrix.
In this specific case, there is only one basis vector $u$, and we can write every possible solution as 
\begin{equation}
\begin{bmatrix} c'_{i_\varrho j_\varrho k_\varrho} \end{bmatrix} 
=
\begin{bmatrix} c'_{i_\varrho j_\varrho k_\varrho} \end{bmatrix}_\text{ref.} 
+  
s \begin{bmatrix} u \end{bmatrix}
\end{equation}
which can be expanded into:
{
\begin{align}
\begin{split}
\begin{bmatrix}
c'_{000}\\
c'_{001}\\
c'_{010}\\
c'_{100}\\
c'_{002}\\
c'_{011}\\
c'_{020}\\
c'_{101}\\
c'_{110}\\
c'_{200}\\
\end{bmatrix}
= 
\varrho_{\text{bulk}}
\left[
\begin{array}{r}
 0.352790 \\
 0.000000 \\
 0.000000 \\
 0.025374 \\
-0.399759 \\
 0.000000 \\
-0.245677 \\
 0.000000 \\
 0.000000 \\
-0.086725 \\
\end{array}
\right]
+
s
\varrho_{\text{bulk}}
\left[
\begin{array}{r}
 0.804494 \\
 0.000000 \\
 0.000000 \\
-0.031540 \\
 0.496907 \\
 0.000000 \\
 0.305381 \\
 0.000000 \\
 0.000000 \\
 0.107801 \\
\end{array}
\right]
\end{split}
\end{align}
}
We can now study the solutions of this sample problem.
First of all, we see that the initial nominal solution can be recovered for $s=0.804494$.
Other solutions are possible, as we show in Figure~\ref{fig:sol_U}, but in order to explore them effectively,
especially when the number of basis vectors increases at higher degrees,
some target functions on the actual mass density values in the body have to be defined.
If we look at the actual mass density generated by the solution,
we can try to find solutions which satisfy additional constraints that can be imposed.
Note how the method described in this manuscript has been free of assumptions up to this point.
The target functions are used to break the degeneracy in the solutions,
as all solutions reproduce exactly the observed gravitational field used as an input.

Before we proceed with defining target functions, 
we need to layout the basic required properties of the solution.
The solution has to be physical, so the minimum mass density has to be greater than zero and not higher than the bulk density,
and the difference between minimum and bulk density, multiplied by the body volume, determines the 
amount of mass which is available to model the interior features of the body.
The degree of the expansion is also important: 
since the number of free parameters in the density distribution 
grows faster than the number of gravity coefficients for a given fixed degree (\S\ref{sec:MatrixTheory}),
then using the same degree for density and gravity generates an under-constrained problem
which is what we are dealing with here, so that's a standard choice.
This also tends to give comparable spatial resolution in density and gravity.

In order to effectively explore the solutions space, 
we need to introduce some assumptions
about the solution, which can be included or excluded as necessary while exploring the solutions space, 
and which translate to a scalar which can be used in a minimum-finding algorithm,
to determine which are the solutions that best satisfy the assumptions.
This approach is similar to what is proposed in \cite{2001Geop...66..511S}.
Examples of target functions are:
\begin{itemize}
\item Density Range (DR): this function enforces the given range in density.
\item Minimum Density Range (MINDR): this target function searches for solutions which have a minimum density range, 
i.e.~are as close as possible to uniform. 
\item Maximum Density Range (MAXDR): similar to the one above, but driving for the largest maximum density range,
which typically results in small regions with very high density, and a large part of the body
close to the minimum density assumed.
\item No Local Minima (NLM): no local minima of the density are allowed except at the boundary (surface) of the body.
\end{itemize}
These target functions are typically implemented by sampling the body's volume with a large number of 
randomly distributed points. Then the mass density is computed at these points, 
and combined into a global scalar target function which we want to minimize.
For the specific cases listed above, explicit target function examples are:
\begin{align}
f_\text{DR}    & = \max(0,\varrho_L-\min(\varrho_k)) + \max(0,\max(\varrho_k)-\varrho_U) \\
f_\text{MINDR} & = \dfrac{\max(\varrho_k)-\min(\varrho_k)}{\varrho_\text{bulk}} \\
f_\text{MAXDR} & = \dfrac{\min(\varrho_k)-\max(\varrho_k)}{\varrho_\text{bulk}} \\
f_\text{NLM}   & = \sum_{k_1 \neq k_2} \max \left( 0, \dfrac{\min(\varrho_{k_1},\varrho_{k_2}) - \varrho_h}{\varrho_\text{bulk}} \right) \dfrac{r_0}{\text{dist}(k_1,k_2)}
\end{align}
where $k,k_1,k_2$ run over the sample points, 
and $\varrho_L$ and $\varrho_U$ are the lower and upper hard limits on the mass density.
The target function $f_\text{NLM}$ compares the density $\varrho_h$, which is computed at the middle point (half-way) $h$ between $k_1$ and $k_2$,
with the densities $\varrho_{k_1}$ and $\varrho_{k_2}$, 
and if the middle density is lower than the two extremes, a positive term (penalty) is included in the sum.
Finally, $\text{dist}(k_1,k_2)$ is the distance between two points.
All these functions are positive when their specific assumption is not met, 
or are zero when the assumption is met,
and as such can be considered penalty functions in the context of a global minimum search.
When several target functions are combined into a single function, their individual weight
needs to be balanced using weighting factors,
which allows us to control directly how the minimum search algorithm will 
try to satisfy different assumptions.
One example is that of having the largest weight for the DR function,
which enforces the solutions to have densities within the lower and upper hard limits,
and then have smaller weight factors for the other functions.

\begin{figure}
\begin{center}
\includegraphics*[bb=215 90 480 440,angle=270,width=0.49\textwidth]{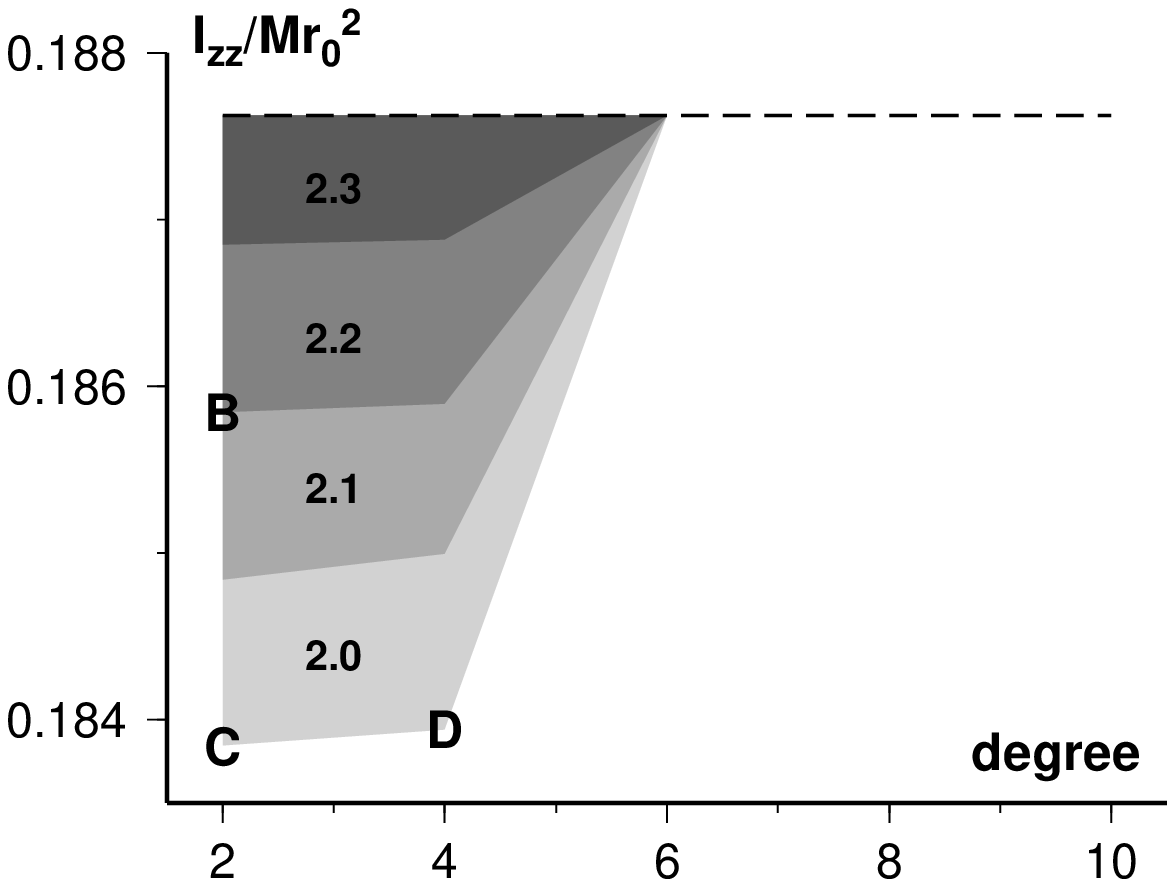}
\end{center}
\caption{
Range of principal inertia moment solutions 
of the sample body in the uniform case,
for a given degree in the gravity and density expansion, 
and for a given value of lowest possible density $\varrho_L$ within the body
as marked by different grey levels.
Note that the range for higher $\varrho_L$ values are fully contained within the range for lower $\varrho_L$ values.
The overall range of inertia moments generated tends to decrease with increasing degree,
converging towards the nominal value of 0.187625 (dashed line) for degree 6 or higher.
The letters B--D mark the values of the corresponding solutions in Figure~\ref{fig:sol_U}.
Only the even degree values have been evaluated in the plot.
}
\label{fig:Izz_range_U}
\end{figure}

In Figure~\ref{fig:sol_U} we present some solutions for degree 2 and 4, for minimum density values of 2.0 and 2.2~g/cm$^3$,
using the target function F=DR+MAXDR+NLM. We note that in all cases where the alternate function F=DR+MINDR+NLM was used,
it always converged to the uniform density solution, and these solutions are not displayed.
In Figure~\ref{fig:Izz_range_U} we show the full range of principal inertia moment solutions, 
obtained using the functions F=DR+MINDR+NLM and F=DR+MAXDR+NLM
to search for high values and low values, respectively.
Note how no solutions other than the nominal one are found for degree 6 or higher.

\subsection{Composite Case}
\label{sec:composite}

\begin{figure}
\begin{center}
\includegraphics*[bb=310 30 550 435,angle=270,width=0.49\textwidth]{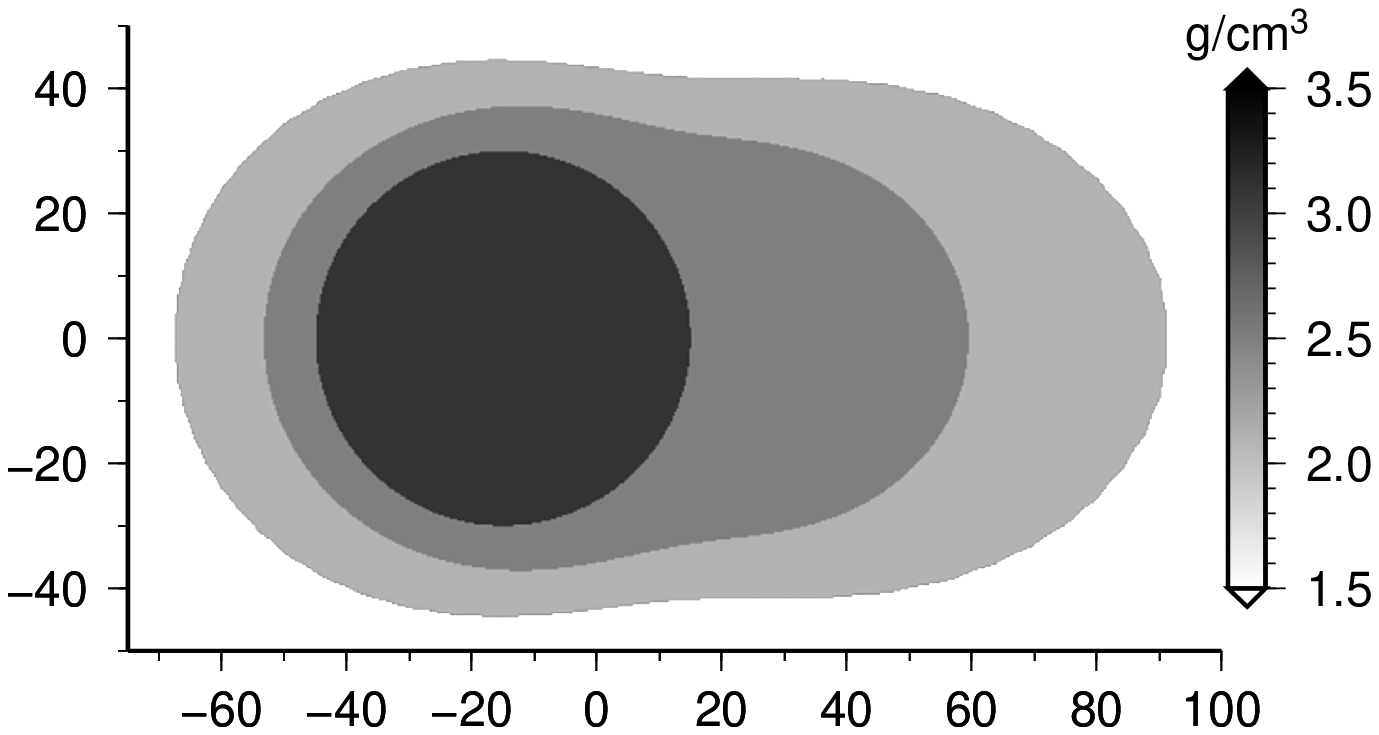}
\end{center}
\caption{Mass distribution of the sample body in the composite case.
The three layers have densities of 2.1, 2.5, and 3.1~g/cm$^3$ moving from the outer to the inner one.
The inner layer is a sphere of 30~km radius offset by -15~km in the $x$ direction, 
while the intermediate layer has a spherical harmonics shape (\S\ref{sec:SMrepSH}) with normalized coefficients of
$\bar{A}_{00}=45$~km,
$\bar{A}_{11}=-4$~km,
$\bar{A}_{20}=-5$~km,
$\bar{A}_{22}=3$~km,
and offset by 10~km in the $x$ direction.
}
\label{fig:B_nominal}
\end{figure}

Now we switch our attention to the layered interior structure for the same shape body,
such as displayed in Figure~\ref{fig:B_nominal}.
In \S\ref{sec:MDrepMAIN} we describe mass distribution
representations which apply to the whole body,
and in \S\ref{sec:SMrepMAIN} we derive an explicit solution for the integral
in Eq.~\eqref{eq:Nijk_expand_generic}
for the most common shape model representations.
Now we analyze the possibility to combine these two results,
to limit a given mass density representation to a give volume within a body,
or to model the interface between two density layers.
Composite models are also necessary because a polynomial description of 
the density is not well suited at modeling sharp density discontinuities:
it can be done, but it requires a very high degree expansion for the density.

Let us consider a body with several components,
each one with specific properties such as total excess mass, shape, position, orientation, and excess density distribution.
The total density function can then be written from Eq.~\eqref{eq:generic_density_expansion} as:
\begin{align}
\varrho(x,y,z) 
= \sum_{\lambda} \varrho_\lambda(x,y,z) 
= \sum_{\lambda} \sum_{i_\varrho j_\varrho k_\varrho=0}^{i_\varrho+j_\varrho+k_\varrho=N_{\varrho,\lambda}} c_{i_\varrho j_\varrho k_\varrho,\lambda} \dfrac{x^{i_\varrho} y^{j_\varrho} z^{k_\varrho}}{r_0^{i_\varrho+j_\varrho+k_\varrho}}
\end{align}
where each component $\lambda$ has an excess mass density $\varrho_\lambda$ within the volume $V_\lambda$,
and $c_{i_\varrho j_\varrho k_\varrho,\lambda}$ are the corresponding coefficients of the power series expansion.
Note that the components can overlap, as long as they are fully contained within the body shape.
The GMoI depend linearly on the density, so we have:
\begin{align}
{\cal N}_{ijk} 
= \sum_{\lambda} {\cal N}_{ijk,\lambda} 
= \dfrac{r_0^3}{M} \sum_{\lambda} \sum_{i_\varrho j_\varrho k_\varrho=0}^{i_\varrho+j_\varrho+k_\varrho=N_{\varrho,\lambda}} c_{i_\varrho j_\varrho k_\varrho,\lambda} \int_{V_\lambda} \dfrac{x^{i+i_\varrho} y^{j+j_\varrho} z^{k+k_\varrho}}{r_0^{i+i_\varrho+j+j_\varrho+k+k_\varrho}} \ \dfrac{{\text d}V}{r_0^3}
\end{align}
Note that when combining the GMoI of components with one of the frequently used 
shape representations considered in \S\ref{sec:SMrepMAIN},
translations and rotations as described in \S\ref{sec:transform} 
might be still necessary in order to express all them in the same reference frame.
Using this we can compute the gravitational field coefficients of Table~\ref{tab:gravity_body_1}
which have to be translated in order to be barycentric, similarly to the uniform case.

\begin{figure}
\begin{center}
\includegraphics*[bb=310 30 550 435,angle=270,width=0.49\textwidth]{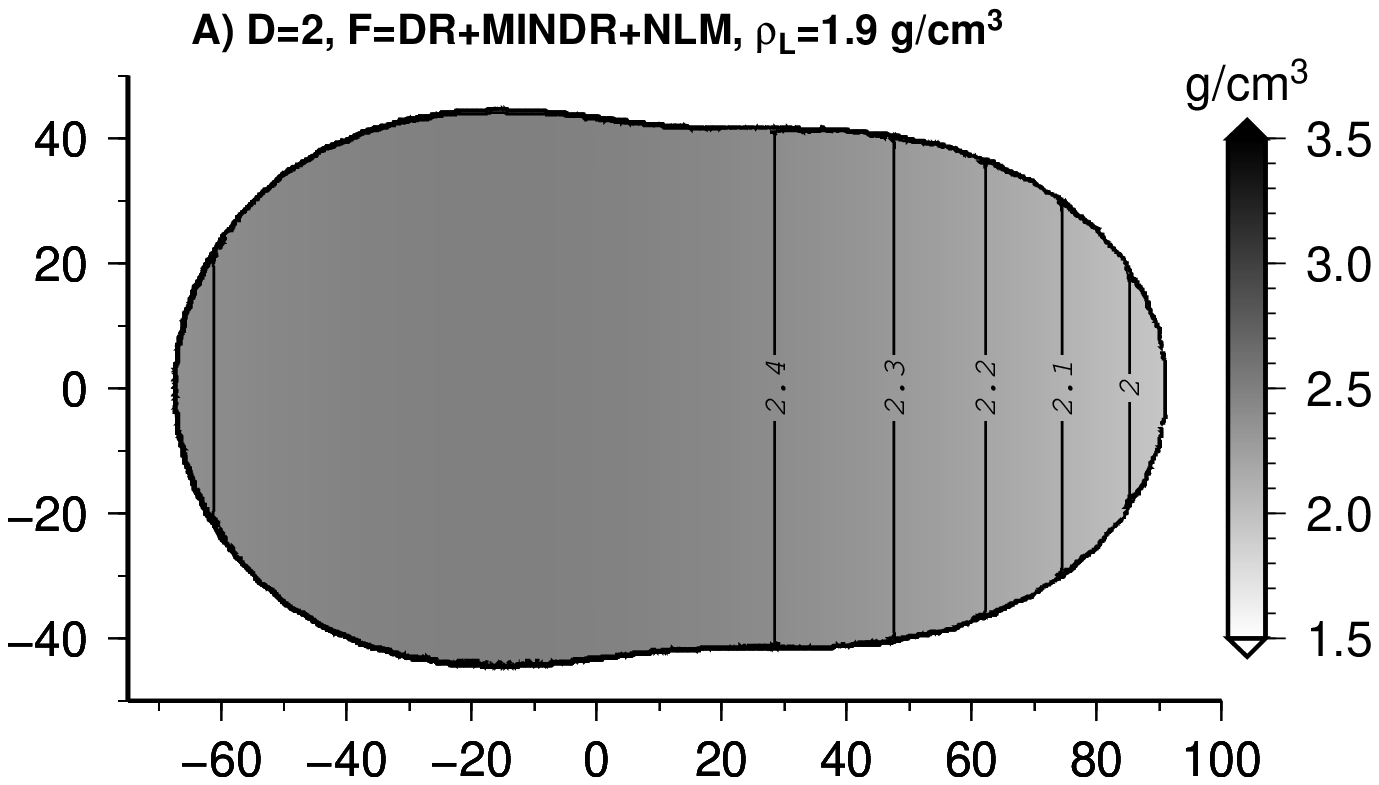}
\includegraphics*[bb=310 30 550 435,angle=270,width=0.49\textwidth]{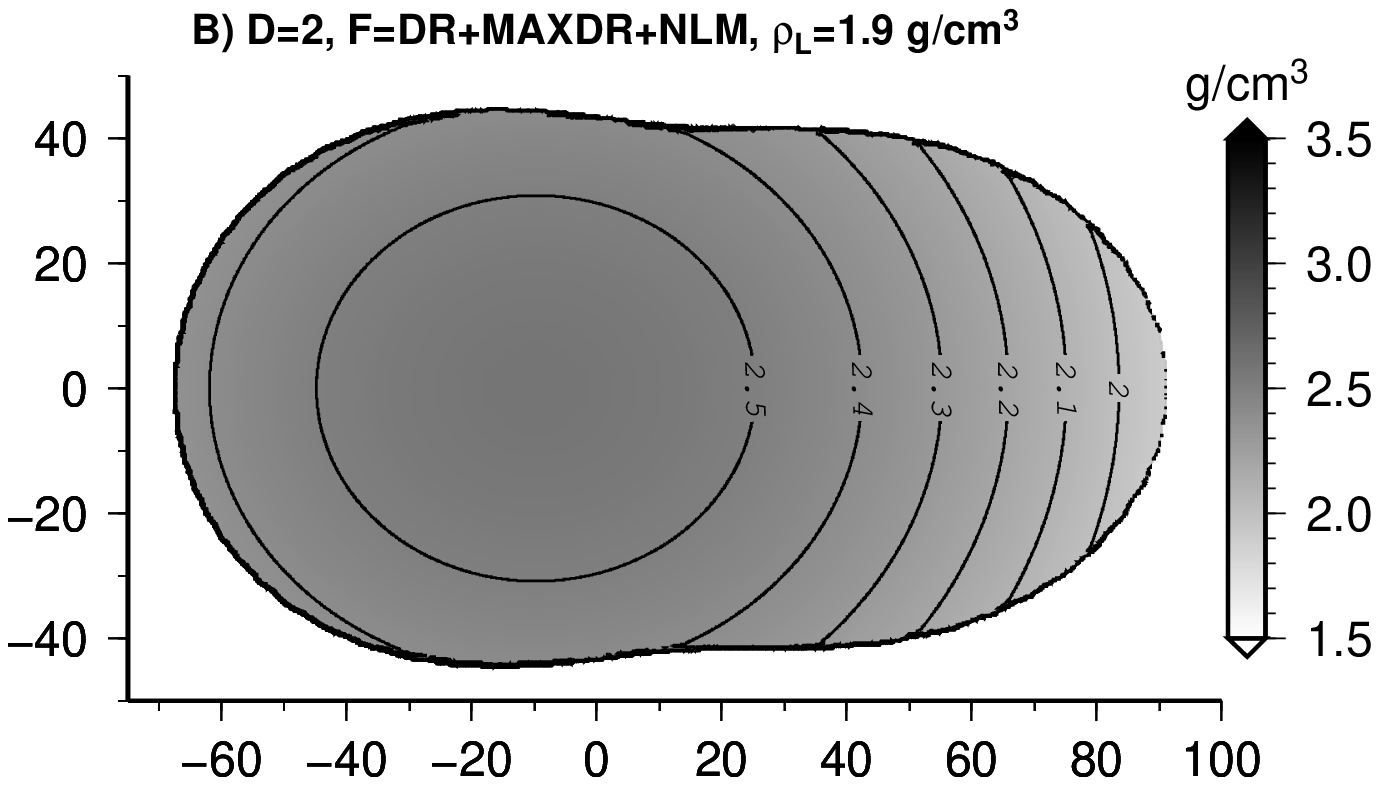}
\includegraphics*[bb=310 30 550 435,angle=270,width=0.49\textwidth]{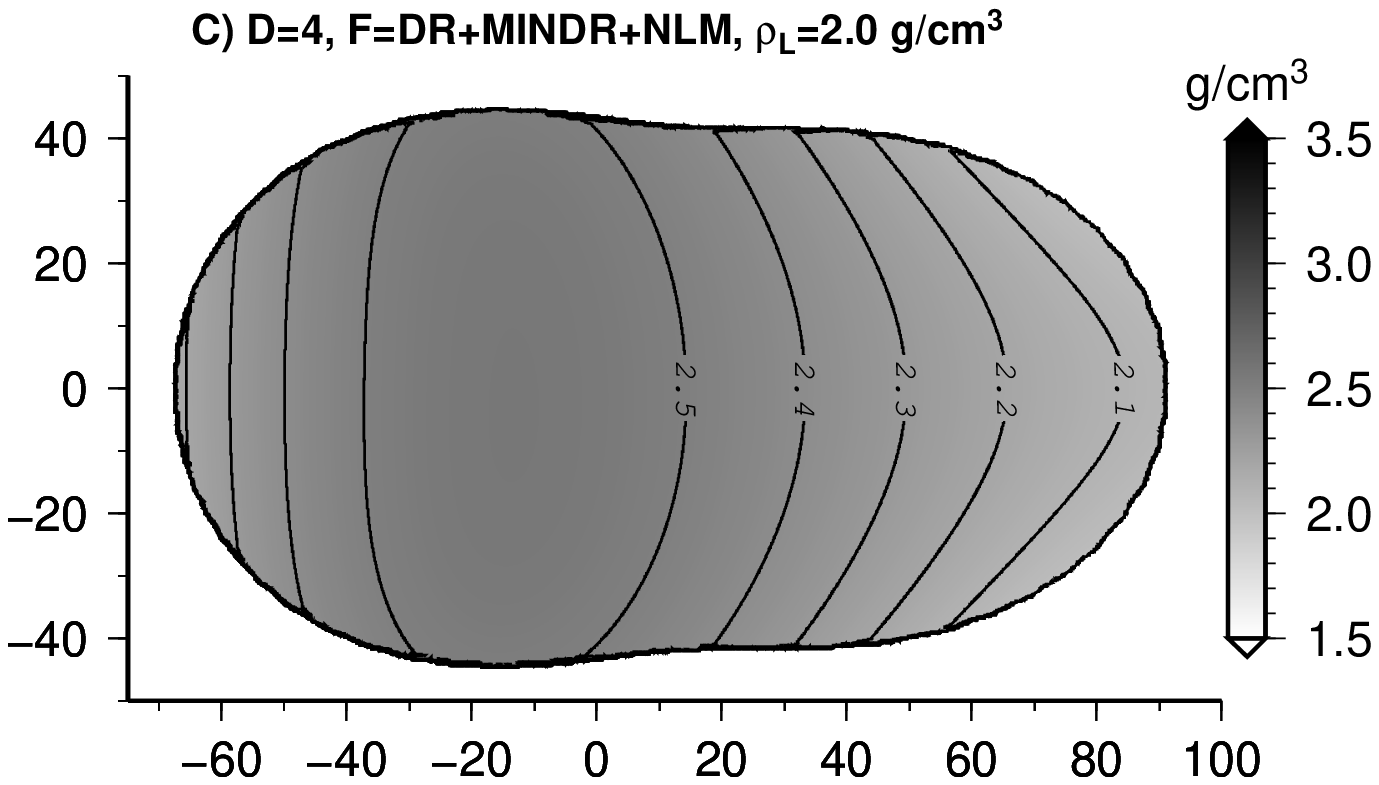}
\includegraphics*[bb=310 30 550 435,angle=270,width=0.49\textwidth]{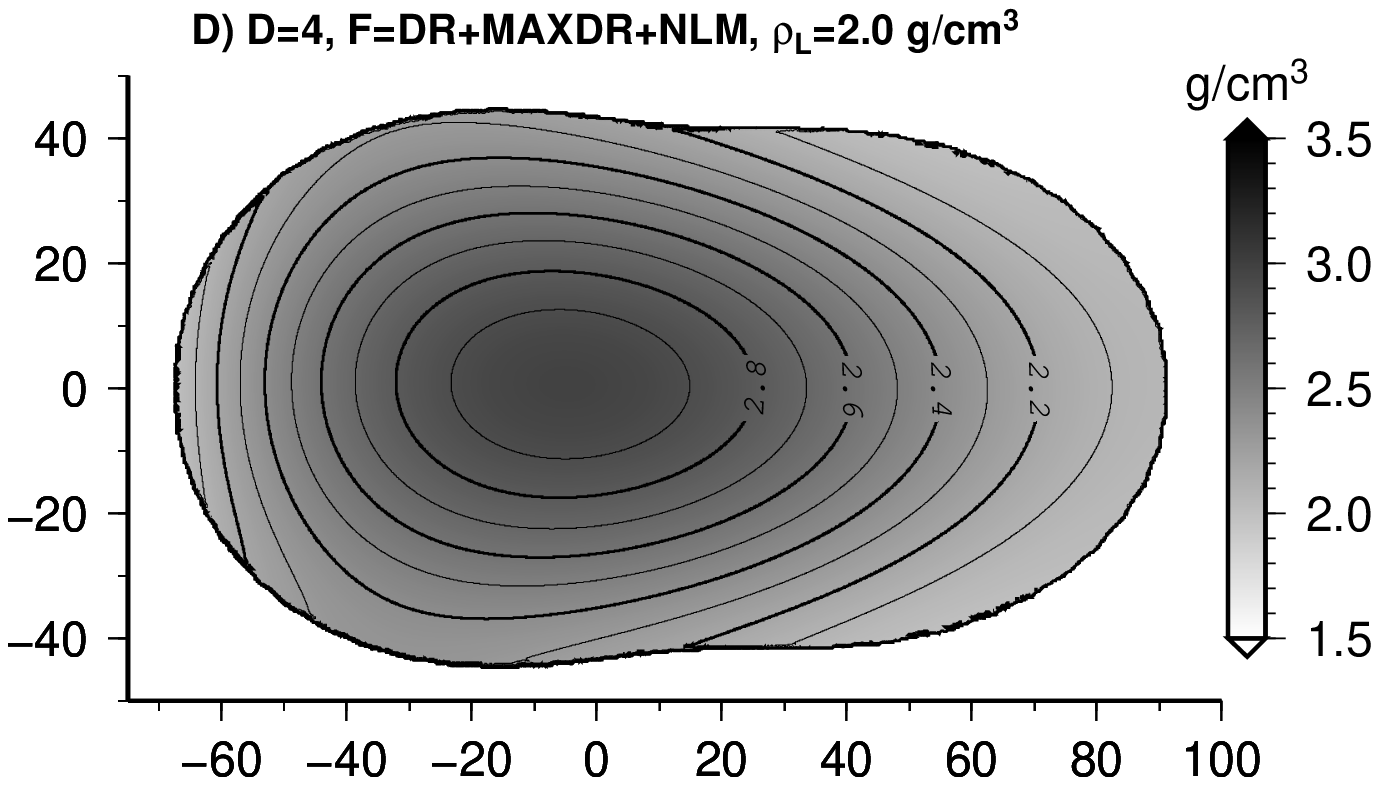}
\includegraphics*[bb=310 30 550 435,angle=270,width=0.49\textwidth]{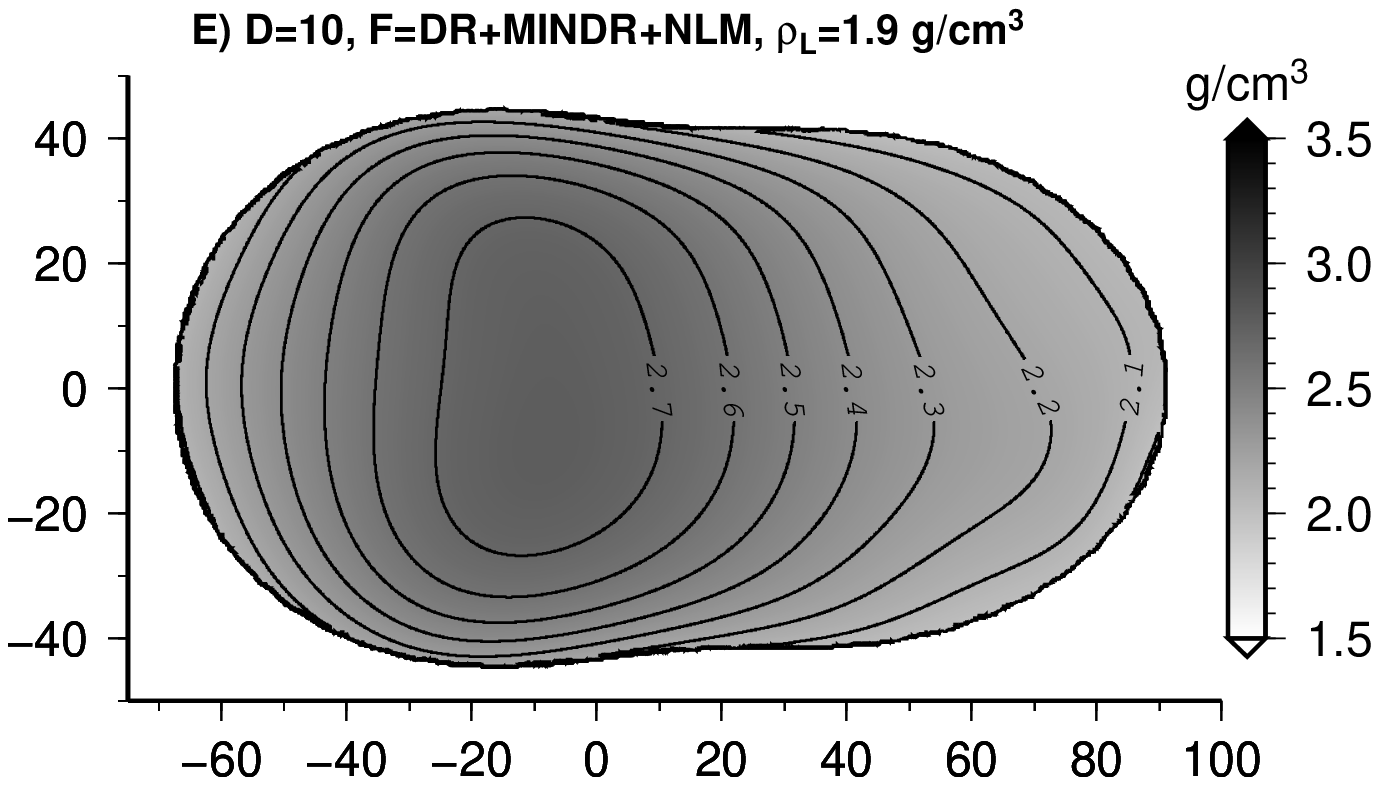}
\includegraphics*[bb=310 30 550 435,angle=270,width=0.49\textwidth]{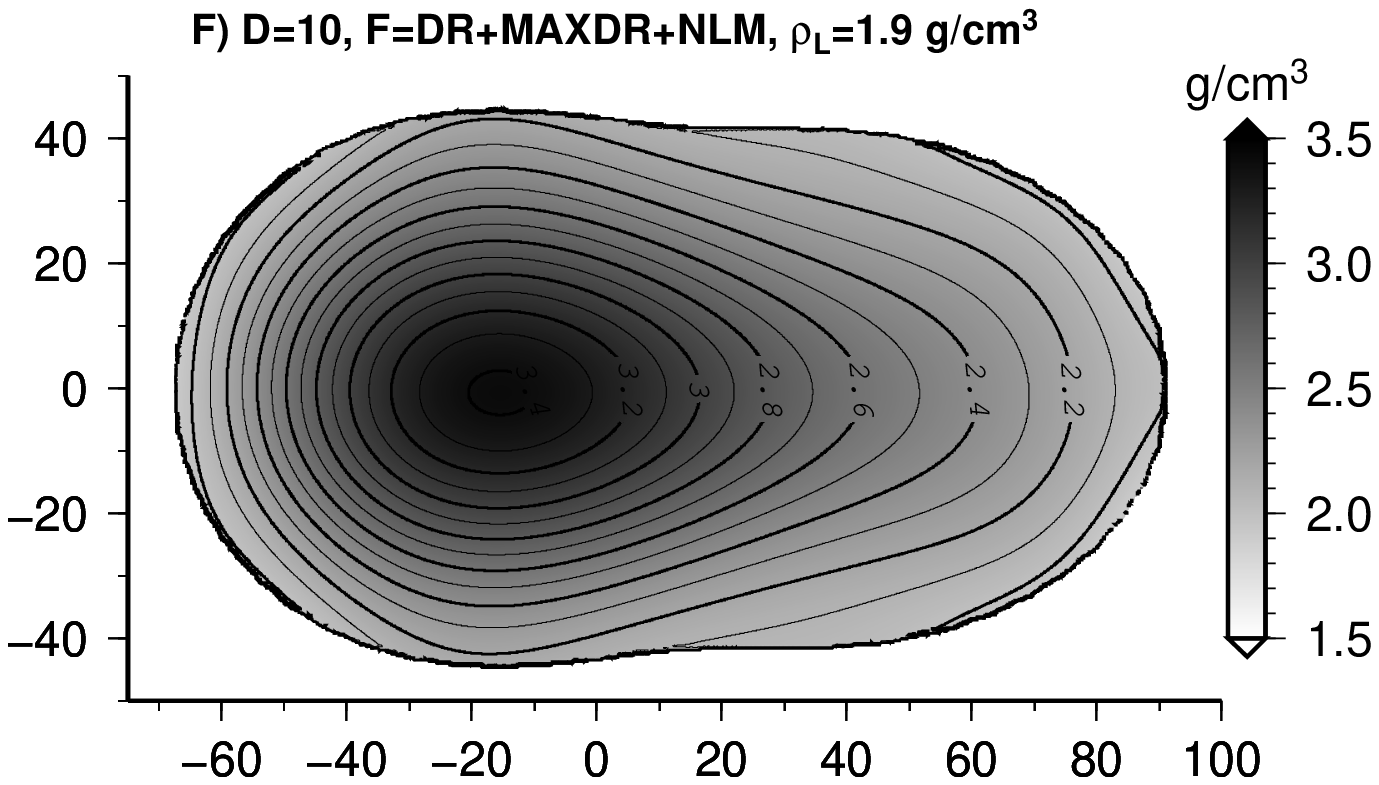}
\end{center}
\caption{Similar to Figure~\ref{fig:sol_U} but for the composite mass distribution case.
Panels (A--F) show different solutions at increasing degree D of gravity and mass distribution,
obtained at a minimum body density $\varrho_L$ value as high as possible to find solutions.
The left column solutions use MINDR to find solutions as close as possible to uniform density,
while the right column solutions use MAXDR to search find large gradients solutions.
Note how the high density region emerges as the degree D increases.
}
\label{fig:sol_B_1}
\end{figure}

\begin{figure}
\begin{center}
\includegraphics*[bb=215 90 480 440,angle=270,width=0.49\textwidth]{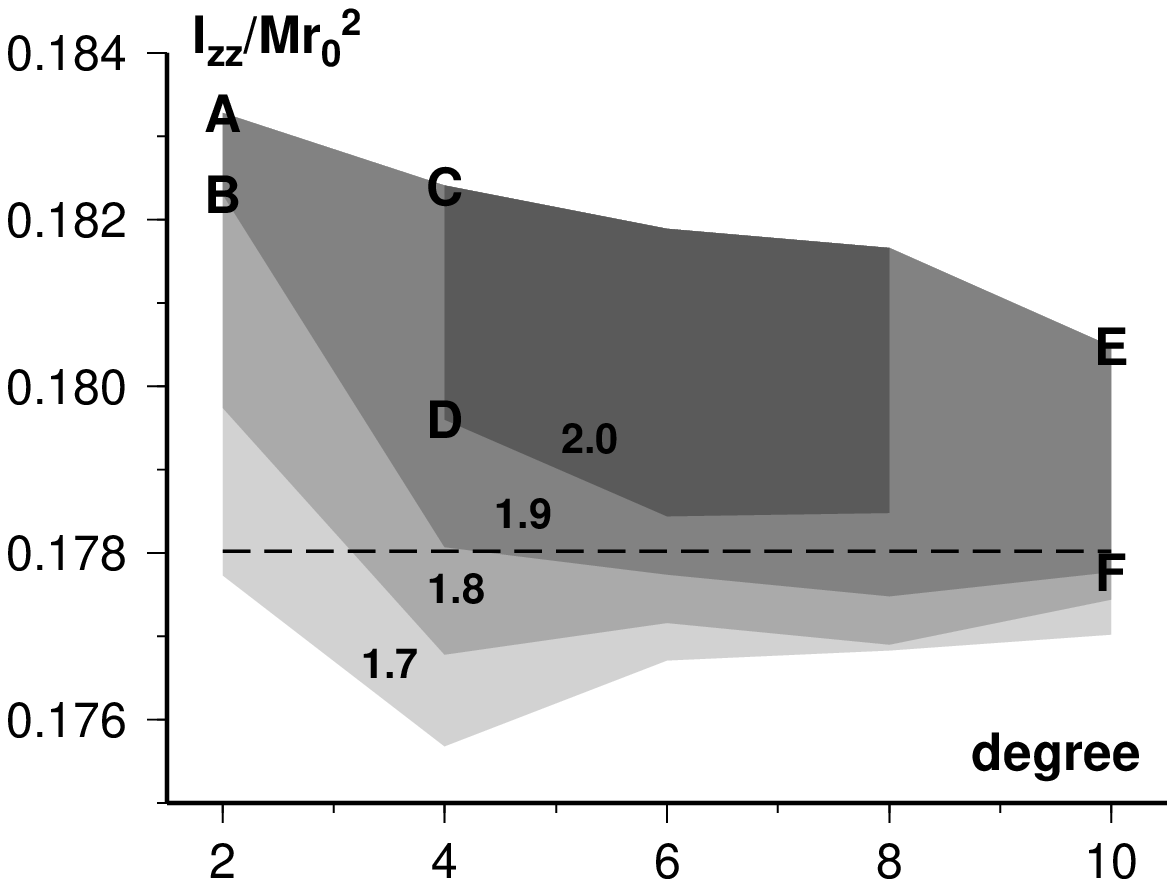}
\end{center}
\caption{
Range of principal inertia moment solutions 
of the sample body in the composite case,
and using the same conventions as in Figure~\ref{fig:Izz_range_U}.
The nominal value of 0.178022 (dashed line)
is now contained within the range of solutions,
and the overall range of inertia moments generated tends to decrease with increasing degree.
Note how the upper end of the range is the same for all $\varrho_L$ values
at a given degree.
The letters A--F mark the values of the corresponding solutions in Figure~\ref{fig:sol_B_1}.
Solutions for $\varrho_L=2.0$~g/cm$^3$ are found only for degree 4 to 8.
}
\label{fig:Izz_range_B}
\end{figure}

Once the gravity coefficients are available, we can solve the inverse problem.
In Figure~\ref{fig:sol_B_1} we present some of the solutions obtained for the composite case,
for varying gravity and mass density degree, target functions, and minimum density value used.
The value of the minimum density is chosen to be as high as possible, at steps of 0.1~g/cm$^3$,
and is the same for each pair of solutions at the same degree D, one using the target 
function F=DR+MINDR+NLM to show the solutions which is closest to uniform density, 
one using F=DR+MAXDR+NLM to show the solution with the largest gradient.
Note how in this case the MINDR solutions are not allowed to become completely uniform,
and their gradient is also forced to increase at increasing degree, 
thus reducing the principal inertia moment (see Figure~\ref{fig:Izz_range_B})

\begin{figure}
\begin{center}
\includegraphics*[bb=310 30 550 435,angle=270,width=0.49\textwidth]{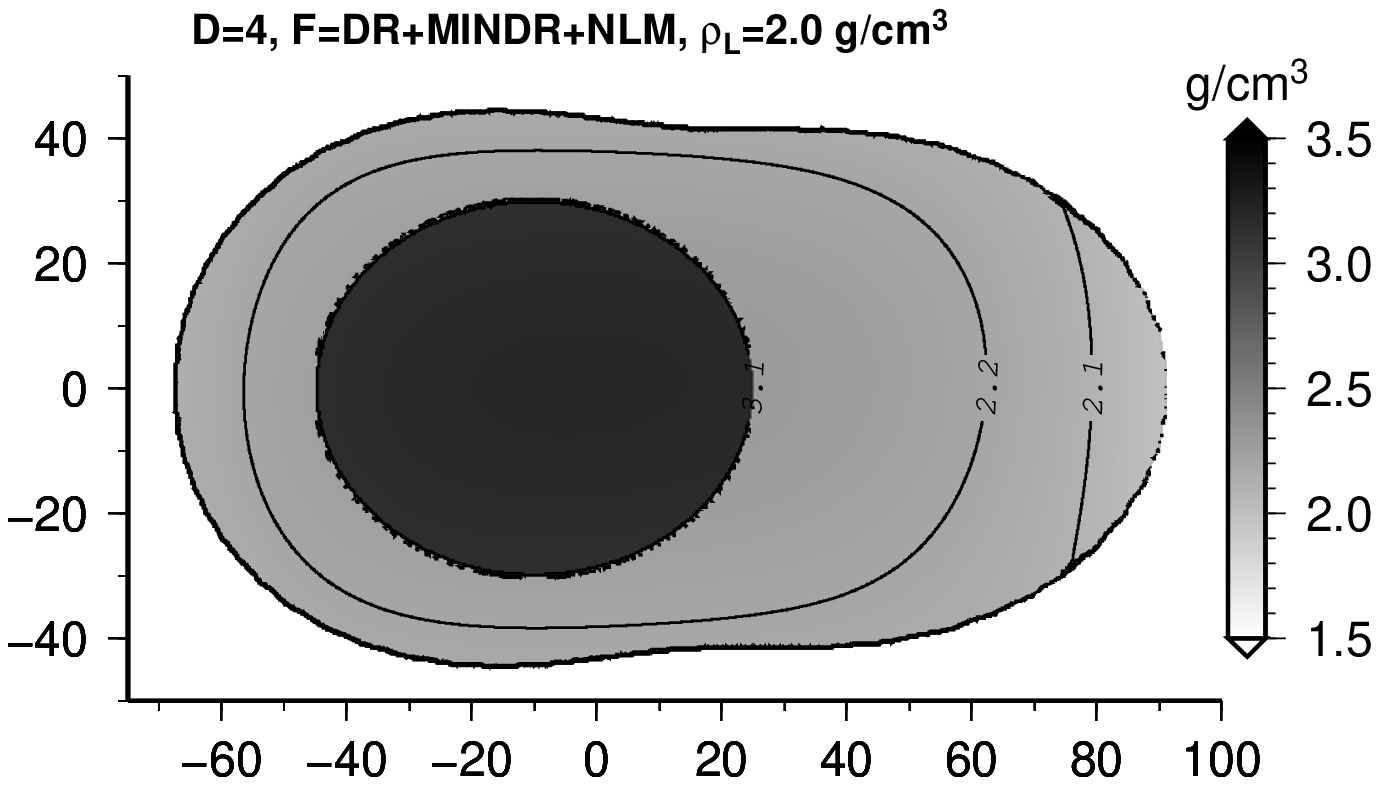}
\end{center}
\caption{Example of composite solution for the composite mass distribution case,
see Figure~\ref{fig:B_nominal} for the nominal distribution.
The ellipsoid has a shape of $35 \times 35 \times 30$~km, 
an excess density of 0.9~g/cm$^3$,
and is offset by -10~km in the $x$ direction.
The inclusion of the ellipsoid allows us to find a global solution where the 
density in the rest of the body is close to uniform.}
\label{fig:sol_B_composite}
\end{figure}

Composite mass distributions can be used also
iteratively during the gravity inversion process.
The idea, as sketched in Figure~\ref{fig:diagram} and also presented more generally in \cite{2006Geop...71J...1S}, 
is to introduce one layer, and then subtract the gravitational field generated 
by it from the observed global field, and then solve the residual field using the standard polynomial approach 
presented earlier.
The global solution can then be used to provide a feedback for the properties of the layer
(shape, position, excess density) in an iterative way, recomputing the gravity generated by the modified layer
and solving for the residual gravity.
One example of the results which can be obtained using this approach
is displayed in Figure~\ref{fig:sol_B_composite},
where an ellipsoid of constant density was included in the solution.
The solution in the figure used the target function F=DR+MINDR+NLM
in order to find solutions which reduce to the minimum the residual gradient,
and its principal inertia moment is $I_{zz} / M r_0^2 = 0.178293$.

\subsection{Realistic Case}
\label{sec:realistic}

In order to investigate the stability of this method in presence of measurement noise,
we add realistic noise to the shape model and separately to the gravitational field, 
using the composite case as the baseline.

For the shape, 
we introduce small random changes to the spherical harmonics coefficients 
in order to obtain a RMS difference $\sigma_s$ with the nominal shape.
Typically the value of $\sigma_s$ depends linearly on the altitude of the mapping orbit and on pixel scales.
For the NEAR mission to the near-Earth asteroid 433~Eros, which has radii ranging from 3.1 to 17.7~km~\citep{2002Icar..155...18T},
a mapping orbit of approximately 35 to 50~km allowed a shape model with RMS error of 3.8~m \citep{2002Icar..155...18T}.
For our sample body with barycentric radii ranging from approximately 44 to 80~km,
a stable mapping orbit would need to be roughly 100~km or larger,
and the shape model resolution $\sigma_s$ should be of the order of 10~m.
The shape is then converted to a discrete triangular mesh with 163,842 vertices,
which is the typical format for irregular bodies.

\begin{figure}
\begin{center}
\includegraphics*[bb=215 90 480 440,angle=270,width=0.49\textwidth]{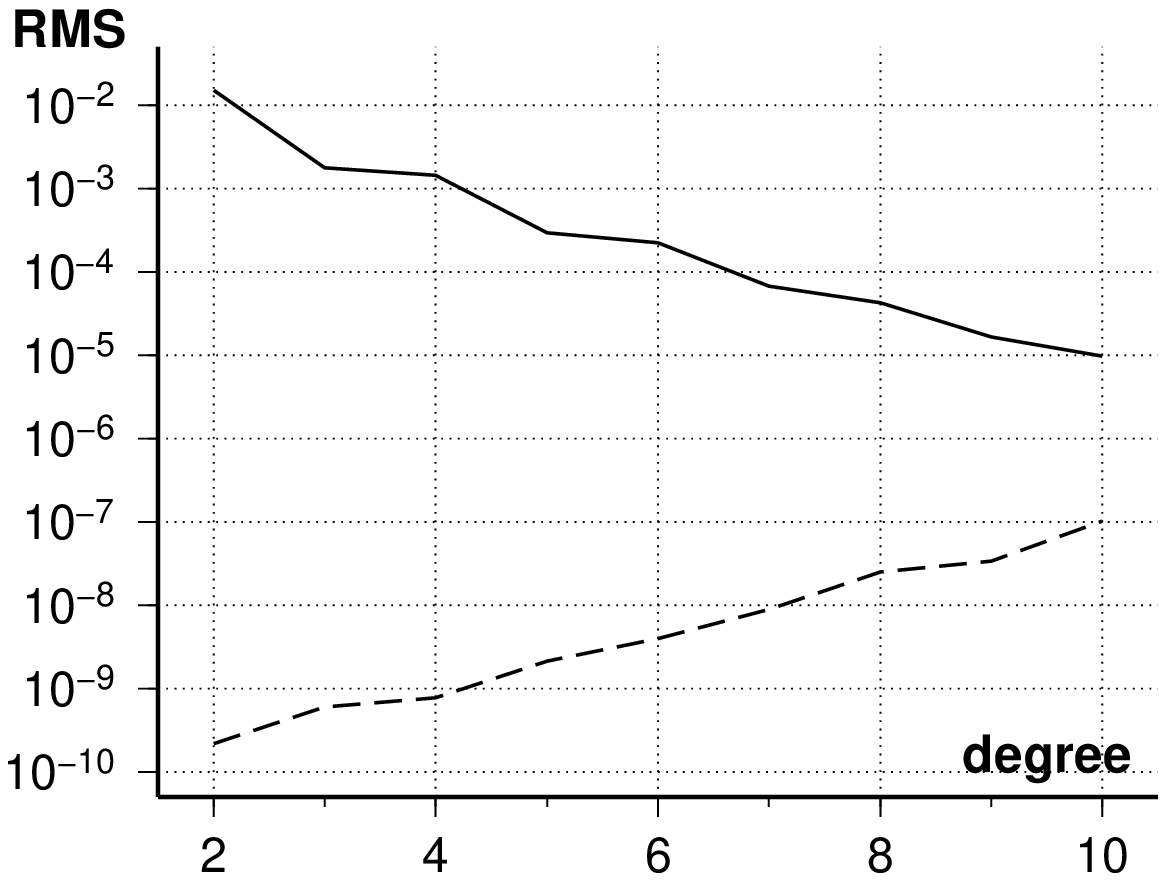}
\end{center}
\caption{RMS magnitude of the gravity signal (solid line) and of the injected noise (dashed line) versus 
the degree of the coefficients, as used in the realistic case.}
\label{fig:RMS_noise}
\end{figure}

For the gravity data, 
the uncertainty typically follows the empirical curve
\begin{align}
\sigma_g (l) \simeq \alpha 10^{\beta (l-l_{\max})} \sqrt{\frac{S_{l_{\max}}^{gg}}{2 l_{\max}+1}}
\end{align}
where 
$S_{l_{\max}}^{gg}$ is the power spectrum of the gravity signal at degree $l_{\max}$,
the slope coefficient $\beta \simeq 1/3$ for 433~Eros \citep{2002Icar..155....3M},
and $\alpha \ll 1$ is the maximum relative magnitude of the noise.
Here we choose $\alpha=0.01$, $\beta=1/3$ and $l_{\max}=10$, see Figure~\ref{fig:RMS_noise}.

\begin{figure}
\begin{center}
\includegraphics*[bb=310 30 550 435,angle=270,width=0.49\textwidth]{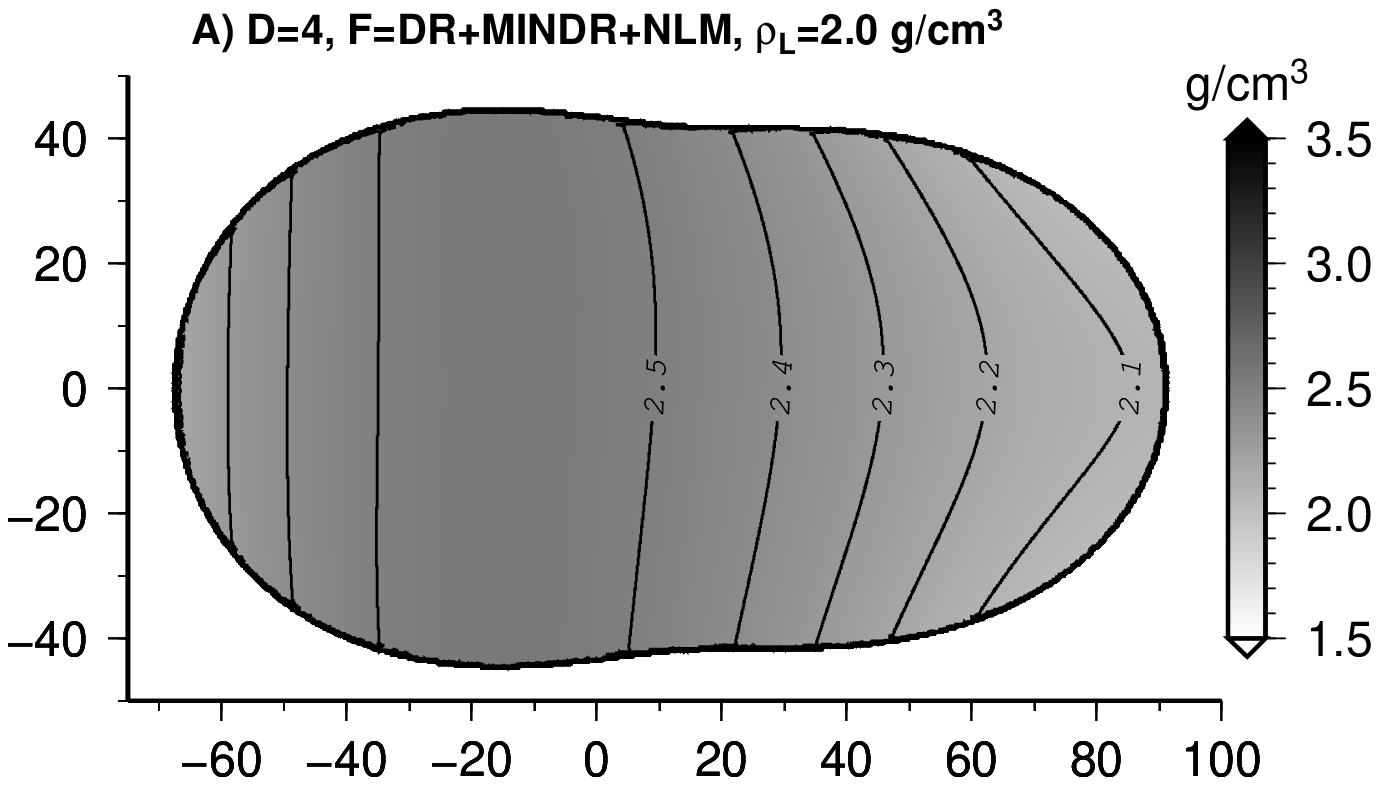}
\includegraphics*[bb=310 30 550 435,angle=270,width=0.49\textwidth]{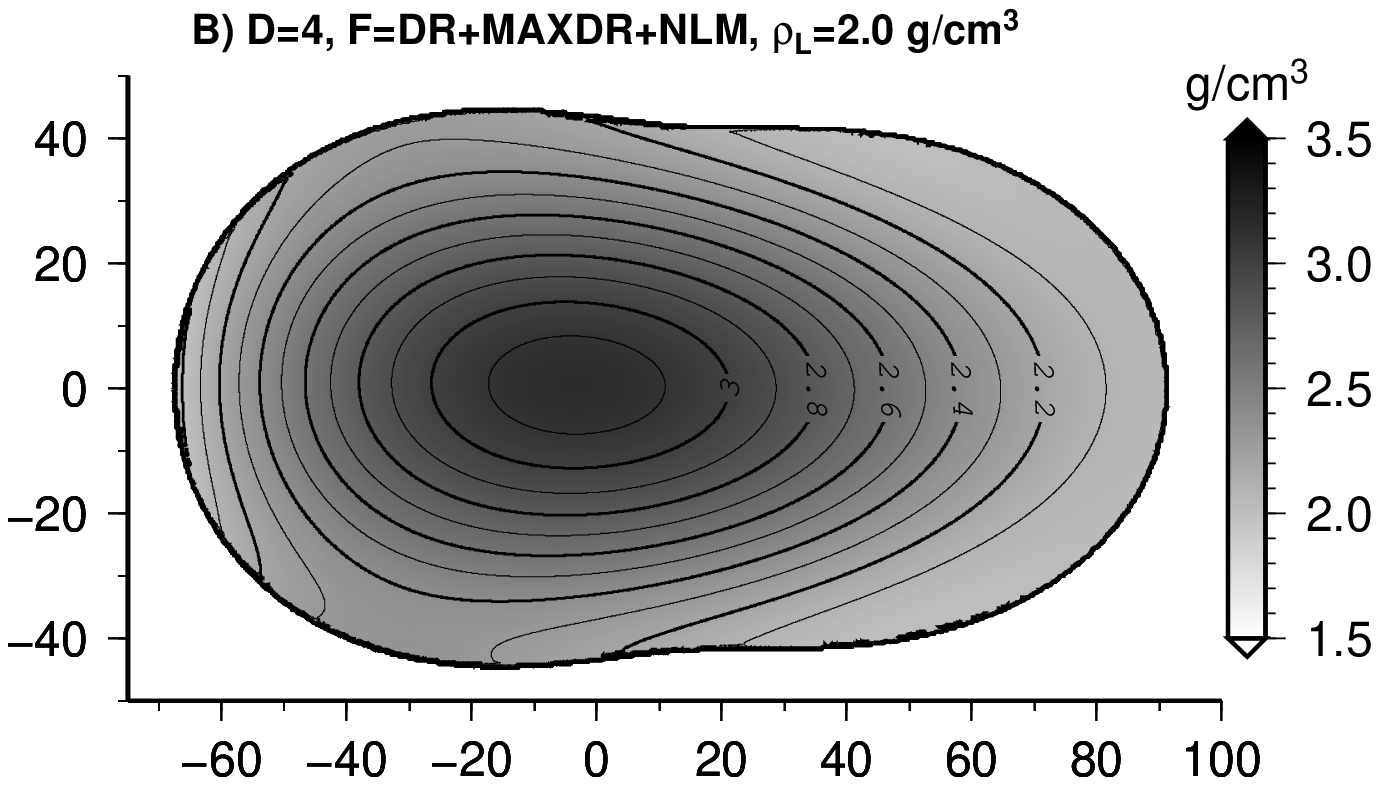}
\includegraphics*[bb=310 30 550 435,angle=270,width=0.49\textwidth]{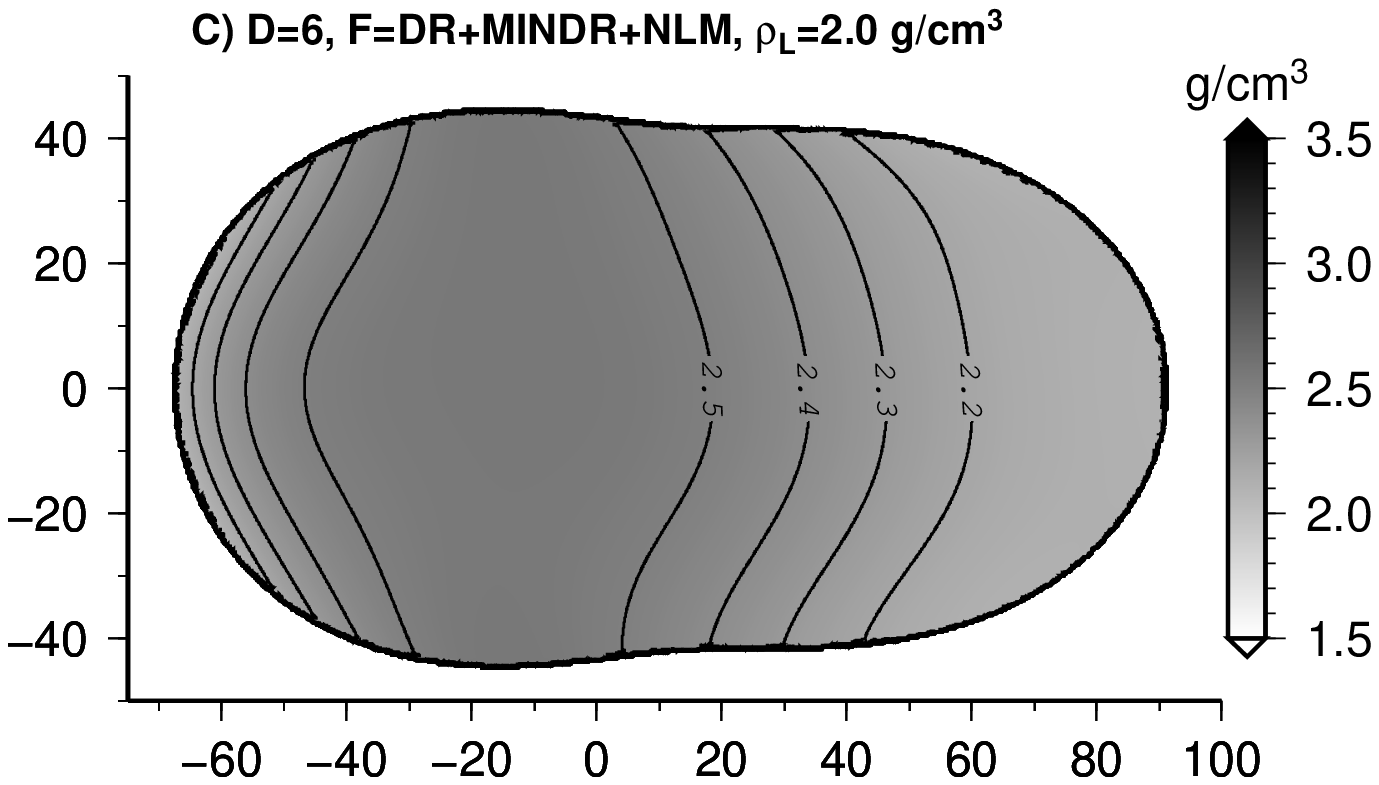}
\includegraphics*[bb=310 30 550 435,angle=270,width=0.49\textwidth]{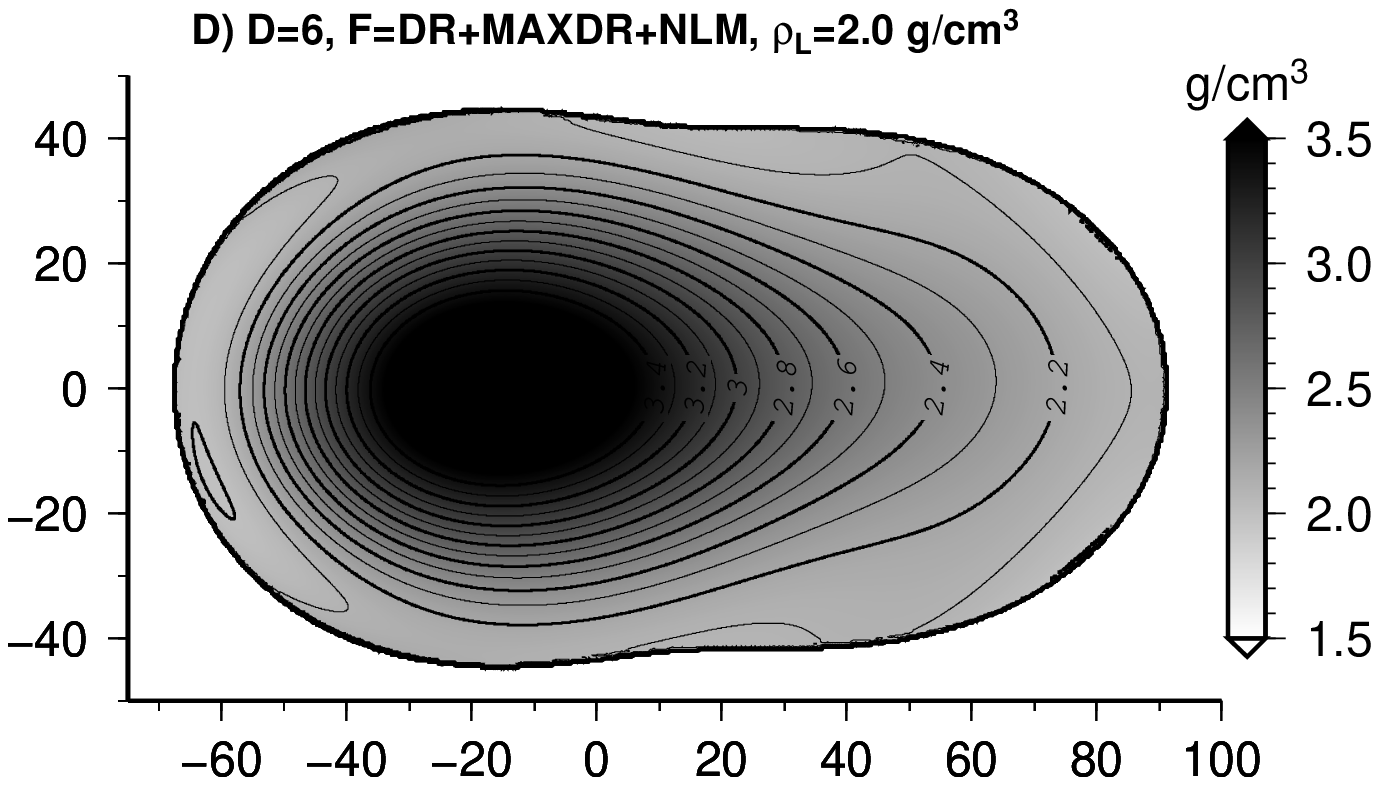}
\includegraphics*[bb=310 30 550 435,angle=270,width=0.49\textwidth]{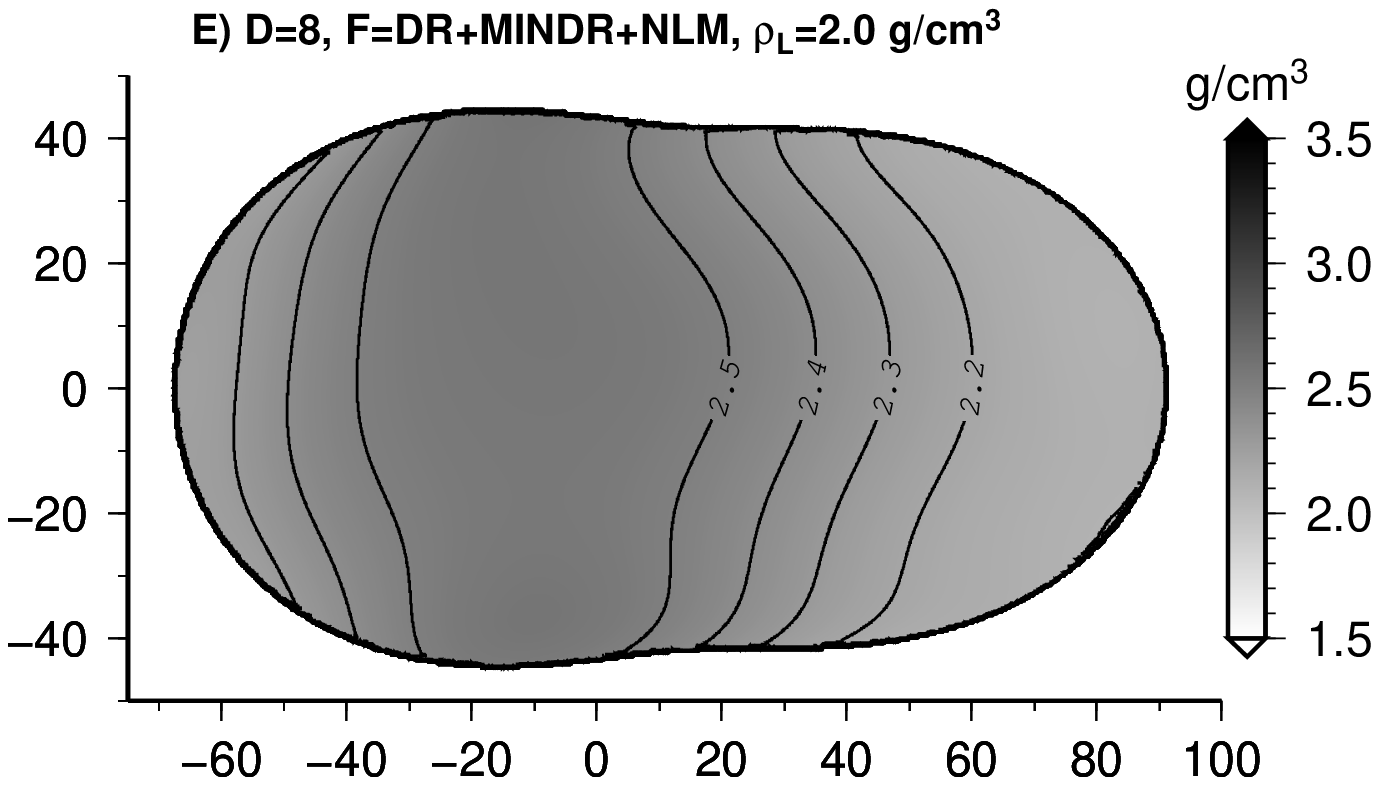}
\includegraphics*[bb=310 30 550 435,angle=270,width=0.49\textwidth]{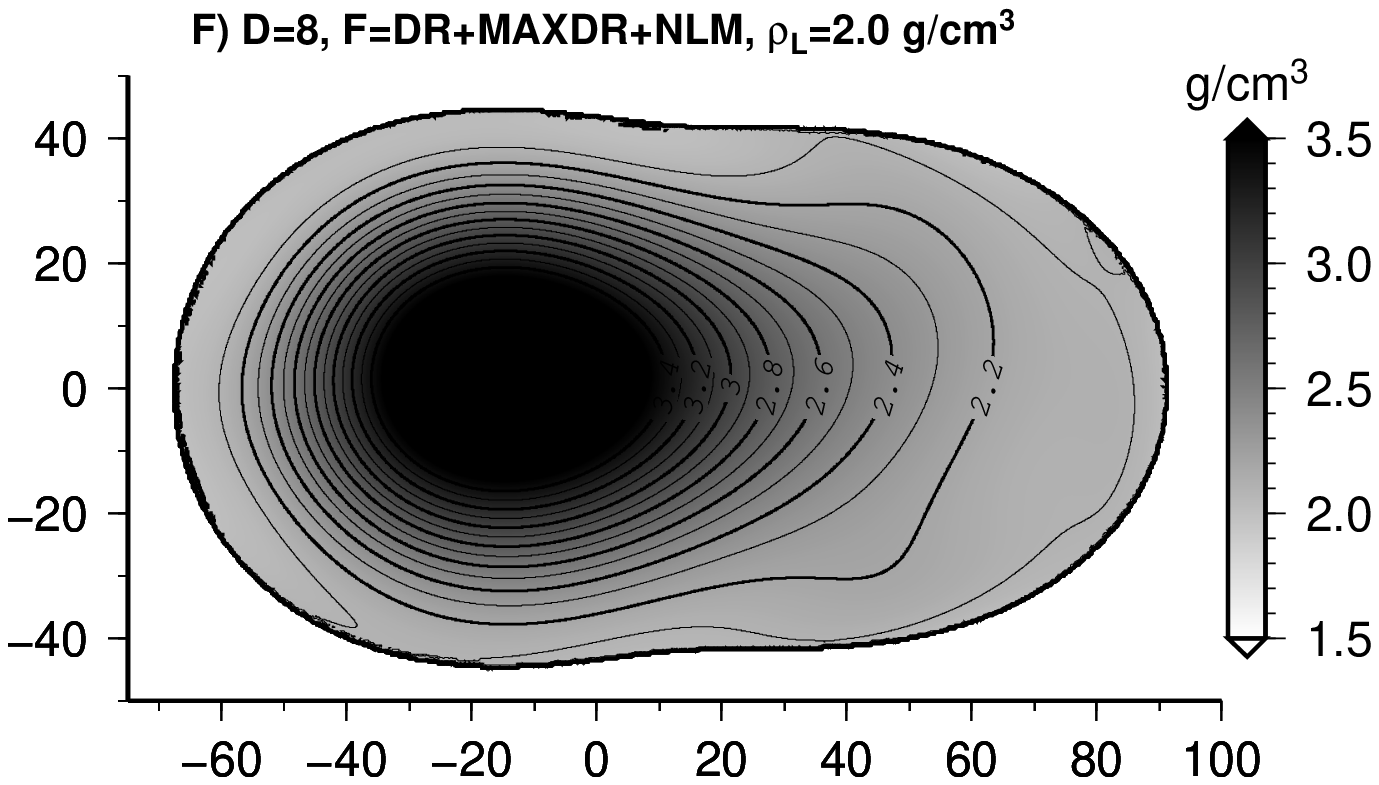}
\end{center}
\caption{Similar to Figure~\ref{fig:sol_U} but for the realistic case.
The degrees D=4, 6, 8 are displayed,
and the minimum density in all cases is 2.0~g/cm$^3$,
while the maximum density reaches 4.1~g/cm$^3$ in panel F.}
\label{fig:sol_noise}
\end{figure}

We find that for the noise levels used here, the shape noise has negligible
effects overall, while the gravity noise becomes important at high degree.
The solutions obtained in presence of noise are displayed in Figure~\ref{fig:sol_noise}.
The degree 4 solutions appear very close to the corresponding noise-free solutions 
in Figure~\ref{fig:sol_B_1}, and at this degree the noise is approximately $10^{-6}$ times the gravity signal.
Solutions are then found for degree 6 and 8 for a noise level up to approximately $10^{-3}$ times the gravity signal, 
and appear to resolve increasingly well the central mass concentration,
even if the central density exceeds significantly the nominal solution.
We were unable to find satisfactory solutions at degree 10, where the noise level is approximately $10^{-2}$ times the gravity signal,
and deduce that this approach might be sensitive to noise of the order of one percent of the signal.
This threshold value might be specific to the case studied,
and so in general it is suggested that solutions are tested at increasing degree
in order to monitor accurately the sensitivity to noise.
The normalized principal inertia moments at degree 8 is between 0.176658 and 0.182300,
thus including the nominal baseline value of 0.178022.

\section{Discussion}
\label{sec:discussion}

The cases studied in \S\ref{sec:sample} outline
what should be the standard approach when using this method:
first, start to search solutions close to uniform with F=DR+MINDR+NLM,
using a minimum density $\varrho_L$ just below the bulk density, 
and at a low degree (2 or 4). If no satisfactory solutions are found,
decrease $\varrho_L$.
Then try F=DR+MAXDR+NLM at the same degree and $\varrho_L$,
in order to obtain solutions with a central high density.
Then, increase the degree and repeat the process.
At some high degree value, the noise will be high enough to 
affect the inversion process and no additional satisfactory solutions will be found.
The solutions obtained can be then interpreted directly,
or used to construct layered models of the interior of the body.

Propagating the uncertainty in the input data (gravity, shape) to the set of interior structure solutions can
be achieved by solving the gravity inversion problem a large number of times,
each time sampling the gravity coefficients from the corresponding covariance matrix, 
and sampling the shape data within its formal uncertainty.
It is possible that the degeneracy in the problem will have a larger effect than
the nominal uncertainty in the input data, so a limited number of test cases will be sufficient to determine this.

The gravity inversion method presented 
has demonstrated a satisfactory behavior in the sample application of \S\ref{sec:sample},
and can be complemented in several ways.
When additional observational constraints are available, these 
can be added in the form of extra equations in 
Eq.~\eqref{eq:compact_CS}.
One such example is the direct observation of a forced precession by the Sun
which can be used to infer the principal inertia moment of the planetary body 
\citep{1973Sci...181..260W,1990JGR....9514137B,2005Icar..175..233B}.

The target functions introduced in \S\ref{sec:sample}
translate assumptions on the solutions
into scalar functions which can be used in the Monte Carlo search.
While the set of target function presented allows us to perform an
initial analysis of the solutions space,
it is possible to develop more complex target functions
which reflect additional assumptions.
Examples include the treatment of the stress tensor associated with a given solution,
or computing its level of isostatic compensation.
A Bayesian approach could also be attempted.
Additionally, we want to stress that every step has been taken in this manuscript
to automate the selection of the solutions presented and to minimize a possible selection bias,
and the target functions provide an excellent solution to this issue by selecting solutions based on the value of a scalar.

In addition to sampling interior structure solutions using a Monte Carlo approach,
an arbitrary density profile can be tested against the solution space by projection.
If we have an arbitrary density distribution 
$\begin{bmatrix} c'_{i_\varrho j_\varrho k_\varrho} \end{bmatrix}_\text{test}$
we can project it over the solutions space using:
\begin{align}
\begin{split}
\begin{bmatrix} c'_{i_\varrho j_\varrho k_\varrho} \end{bmatrix}_\text{proj.} 
& =
\begin{bmatrix} c_{i_\varrho j_\varrho k_\varrho} \end{bmatrix}_\text{ref.} 
+
\sum_q
s_q
\begin{bmatrix} u_q \end{bmatrix} \\
s_q 
& = 
\left(
\begin{bmatrix} c'_{i_\varrho j_\varrho k_\varrho} \end{bmatrix}_\text{test} -
\begin{bmatrix} c'_{i_\varrho j_\varrho k_\varrho} \end{bmatrix}_\text{ref.} 
\right)
\cdot
\begin{bmatrix} u_q \end{bmatrix}
\end{split}
\end{align}
where the $\cdot$ is the scalar product between the two vectors.
In the case of a composite solution, 
we first need to subtract for the gravity of the constant components,
and then solve for the residual gravitational coefficients to
obtain a residual reference solution.
An example of a composite solution was provided in Figure~\ref{fig:sol_B_composite},
where an ellipsoid with uniform density was included in the solution,
and then an exact global solution was obtained on the residual gravity.

Working with layers with variable characteristics is possible, 
and a spherical harmonics representation of the surface (see \S\ref{sec:SMrepSH}) can prove very flexible in this context,
but the GMoI are not linear in $A_{lm}$ e $B_{lm}$, see Eq.~\eqref{eq:rn_final},
so the shape cannot be found directly by linear inversion.
This issue can be resolved in several ways, including iterative methods,
where initial values for the shape coefficients $A_{lm}$ e $B_{lm}$ and for the excess density are guessed,
then the gravity generated by this layer is computed,
and the model is finally solved using the standard method presented here.
This global solution can then be used to provide a feedback on the parameters of the layer (see Figure~\ref{fig:diagram}), 
to modify them and then repeat the same steps with the updated layer parameters.
An alternative independent approach is also provided by spectral methods 
\citep{1973GeoJI..31..447P,1977AREPS...5...35P,1998JGR...103.1715W},
where a candidate solution can be used as input 
in either an iteration or a projection scheme.

The results in \S\ref{sec:SMrepSH} allow us to derive the inertial and gravitational properties of a body with
arbitrary mass density distribution and with an exterior shape described using spherical harmonics,
improving over \cite{1994CeMDA..60..331B} where similar results were obtained for homogeneous bodies only.

\section{Conclusions}

We have combined several analytical tools to obtain a gravity inversion method
which generates exact solutions for a planetary body with a given shape, rotation, and gravity.
Orthogonal polynomials are used to expand the mass density function within the body,
and this allows us to obtain a linear map between density and global gravity.
This map is then pseudo-inverted in the under-constrained regime,
and QR decomposition provides a basis of the non-trivial null space
of all the degenerate interior solutions which produce the observed gravity.
In order to break degeneracy, assumptions are introduced,
which can be transformed into scalar target functions on the mass density distribution,
and a Monte Carlo approach can be used to explore the solutions space while minimizing these functions,
to satisfy the corresponding assumptions.
Layers can also be included in the model, 
with shape and density which can be modified iteratively.

Sample applications show that as long as the assumptions are correct,
the solutions generated tend to converge towards the nominal interior structure of a planetary body.
This is confirmed by an inspection of the sections of the sample body,
and also by monitoring the range of principal inertia moments generated by the solutions.
Solutions are stable in presence of moderate noise, 
but this can limit the highest degree which produces satisfactory solutions.

The mathematical formalism is presented in great detail,
and the material in the Appendix sections should make this approach
immediately applicable to a wide range of problems.

\begin{acknowledgments}
We gratefully acknowledge two anonymous referees who contributed to improving this 
manuscript with their comments.
This research was supported by the NASA DAVPS program, grant NNX10AR20G,
and made use of NASA's Astrophysics Data System.
This document was prepared using the \LaTeX{} typesetting system,
and figures \ref{fig:sol_U}--\ref{fig:sol_noise} were generated using the Generic Mapping Tools \citep{1991EOSTr..72..441W}.
\end{acknowledgments}

\appendix

\section{Transformations of the Generalized Moments of Inertia}
\label{sec:transform}

Under translation of the coordinates from $(x,y,z)$ to $(x',y',z')=(x-\Delta_x,y-\Delta_y,z-\Delta_z)$,
we have that the GMoI as defined in Eq.~\eqref{eq:Nijk_expand_generic} transform as:
\begin{align}
\label{eq:Nijk_translation} 
\begin{split}
{\cal N}'_{ijk} 
= & \dfrac{r_0^3}{M} \sum_{i_\varrho j_\varrho k_\varrho=0}^{i_\varrho+j_\varrho+k_\varrho=N_\varrho} c_{i_\varrho j_\varrho k_\varrho} \int_{V} \dfrac{(x-\Delta_x)^{i} (y-\Delta_y)^{j} (z-\Delta_z)^{k}}{r_0^{i+j+k}} \dfrac{x^{i_\varrho} y^{j_\varrho} z^{k_\varrho}}{r_0^{i_\varrho+j_\varrho+k_\varrho}} \ \dfrac{{\text d}V}{r_0^3} \\
= & \dfrac{r_0^3}{M} \sum_{i_\varrho j_\varrho k_\varrho=0}^{i_\varrho+j_\varrho+k_\varrho=N_\varrho} c_{i_\varrho j_\varrho k_\varrho} \sum_{b_i=0}^{i} \sum_{b_j=0}^{j} \sum_{b_k=0}^{k} (-1)^{b_i+b_j+b_k} \binom{i}{b_i} \binom{j}{b_j} \binom{k}{b_k} \\ 
  & \dfrac{\displaystyle \Delta_x^{b_i} \Delta_y^{b_j} \Delta_z^{b_k}}{r_0^{b_i+b_j+b_k}} \int_{V} \dfrac{x^{i+i_\varrho-b_i} y^{j+j_\varrho-b_j} z^{k+k_\varrho-b_k}}{r_0^{i+i_\varrho-b_i+j+j_\varrho-b_j+k+k_\varrho-b_k}} \ \dfrac{{\text d}V}{r_0^3} \\
= & \sum_{b_i=0}^{i} \sum_{b_j=0}^{j} \sum_{b_k=0}^{k} (-1)^{b_i+b_j+b_k} \binom{i}{b_i} \binom{j}{b_j} \binom{k}{b_k} \dfrac{\displaystyle \Delta_x^{b_i} \Delta_y^{b_j} \Delta_z^{b_k}}{r_0^{b_i+b_j+b_k}} {\cal N}_{i-b_i,j-b_j,k-b_k} 
\end{split}
\end{align}
The transformation relation for rotations can also be obtained in a similar fashion.
If $m_{ij}$ are the elements of the matrix $M$ transforming $(x,y,z)$ to $(x',y',z')$,
then we have:
\begin{align}
\label{eq:Nijk_rotation} 
\begin{split}
{\cal N}'_{ijk} 
= & \dfrac{r_0^3}{M} \sum_{i_\varrho j_\varrho k_\varrho=0}^{i_\varrho+j_\varrho+k_\varrho=N_\varrho} c_{i_\varrho j_\varrho k_\varrho} \\
  & \int_{V} \dfrac{(m_{11}x+m_{12}y+m_{13}z)^{i} (m_{21}x+m_{22}y+m_{23}z)^{j} (m_{31}x+m_{32}y+m_{33}z)^{k}}{r_0^{i+j+k}} \dfrac{x^{i_\varrho} y^{j_\varrho} z^{k_\varrho}}{r_0^{i_\varrho+j_\varrho+k_\varrho}} \ \dfrac{{\text d}V}{r_0^3} \\
= & \dfrac{r_0^3}{M} \sum_{i_\varrho j_\varrho k_\varrho=0}^{i_\varrho+j_\varrho+k_\varrho=N_\varrho} c_{i_\varrho j_\varrho k_\varrho} \sum_{\substack{s_{11}+s_{12}+s_{13}=i \\ s_{21}+s_{22}+s_{23}=j \\ s_{31}+s_{32}+s_{33}=k}} \dfrac{i!}{s_{11}!s_{12}!s_{13}!} \dfrac{j!}{s_{21}!s_{22}!s_{23}!} \dfrac{k!}{s_{31}!s_{32}!s_{33}!} \prod_{p,q=1}^{3} m_{pq}^{s_{pq}} \\
  & \int_{V} \dfrac{x^{s_{11}+s_{21}+s_{31}+i_\varrho} y^{s_{12}+s_{22}+s_{32}+j_\varrho} z^{s_{13}+s_{23}+s_{33}+k_\varrho}}{r_0^{i+i_\varrho+j+j_\varrho+k+k_\varrho}} \ \dfrac{{\text d}V}{r_0^3} \\
= & \sum_{\substack{s_{11}+s_{12}+s_{13}=i \\ s_{21}+s_{22}+s_{23}=j \\ s_{31}+s_{32}+s_{33}=k}} \dfrac{i!}{s_{11}!s_{12}!s_{13}!} \dfrac{j!}{s_{21}!s_{22}!s_{23}!} \dfrac{k!}{s_{31}!s_{32}!s_{33}!} \prod_{p,q=1}^{3} m_{pq}^{s_{pq}} {\cal N}_{s_{11}+s_{21}+s_{31},s_{12}+s_{22}+s_{32},s_{13}+s_{23}+s_{33}} 
\end{split}
\end{align}

\section{Mass Density Representation}
\label{sec:MDrepMAIN}

The particular choice of orthogonal polynomials $P_{w_1 w_2 \cdots}(x,y,z)$ used as a basis
for the expansion of the mass density function (see Eq.~\eqref{eq:generic_density_expansion})
determines the explicit expression of the GMoI.
In the trivial case of a power series expansion, the polynomials are identically $P_{w_1 w_2 \cdots}(x,y,z)=1$ for all $w_1 w_2 \cdots$,
and the expressions for Eq.~\eqref{eq:generic_density_expansion} and Eq.~\eqref{eq:Nijk_expand_generic} are unchanged.

If we consider instead the Chebyshev polynomials for a basis, we have:
\begin{align}
  \varrho(x,y,z) & = \sum_{i_\varrho j_\varrho k_\varrho=0}^{i_\varrho+j_\varrho+k_\varrho=N_\varrho} c'_{i_\varrho j_\varrho k_\varrho} T_{i_\varrho}(x/r_0) T_{j_\varrho}(y/r_0) T_{k_\varrho}(z/r_0)
\end{align}
where $T_n(x)$ are Chebyshev polynomials of the first kind of degree $n$.
Chebyshev polynomials are defined by the recurrence relation 
$T_n(x) = 2 x T_{n-1}(x) - T_{n-2}(x)$, with $T_0(x)=1$ and $T_1(x)=x$,
so if we express this as 
$T_n(x) = \sum_{m=0}^n t(n,m) x^m$
we can obtain the following recursive relation for the coefficient $t(n,m)$:
\begin{align}
t(n,m) = 2 t(n-1,m-1) - t(n-2,m)
\end{align}
where $0 \leq m \leq n$
and the first terms are 
$t(0,0) = 1$,
$t(1,0) = 0$,
$t(1,1) = 1$.
The corresponding expression for the mass density function is then:
\begin{align}
\varrho(x,y,z) & = \sum_{i_\varrho j_\varrho k_\varrho=0}^{i_\varrho+j_\varrho+k_\varrho=N_\varrho} \sum_{s_i=0}^{i_\varrho} \sum_{s_j=0}^{j_\varrho} \sum_{s_k=0}^{k_\varrho} c'_{s_i s_j s_k} t(i_\varrho,s_i) t(j_\varrho,s_j) t(k_\varrho,s_k) \dfrac{x^{i_\varrho} y^{j_\varrho} z^{k_\varrho}}{r_0^{i_\varrho+j_\varrho+k_\varrho}} \\
c_{i_\varrho j_\varrho k_\varrho} & = \sum_{s_i=0}^{i_\varrho} \sum_{s_j=0}^{j_\varrho} \sum_{s_k=0}^{k_\varrho} c'_{s_i s_j s_k} t(i_\varrho,s_i) t(j_\varrho,s_j) t(k_\varrho,s_k)
\end{align}
which can be used directly in Eq.~\eqref{eq:Nijk_expand_generic}.
Explicit expressions for different choices of orthogonal polynomials can be obtained in a similar fashion.

\section{Shape Model Representation}
\label{sec:SMrepMAIN}

The computation of the GMoI using Eq.~\eqref{eq:Nijk_expand_generic} depends on the volume integral in Eq.~\eqref{eq:integral_def}:
\begin{align}
\Phi_{i+i_\varrho,j+j_\varrho,k+k_\varrho}(r_0,V) & = \int_{V} \dfrac{x^{i+i_\varrho} y^{j+j_\varrho} z^{k+k_\varrho}}{r_0^{i+i_\varrho+j+j_\varrho+k+k_\varrho}} \ \dfrac{{\text d}V}{r_0^3}
\end{align}
over the body volume. 
In this section we solve this volume integral
when the shape of the body is provided in one of several common representations.

\subsection{Triaxial Ellipsoid}
\label{sec:SMrepEll}

The triaxial ellipsoid is one of the simplest regular shape to approximate natural bodies.
The GMoI of an uniform ellipsoid with semi-axes $a,b,c$ is \citep{2008CeMDA.100..319T}:
\begin{align}
\begin{split}
{\cal N}_{ijk} 
& = 3 \frac{\displaystyle a^i b^j c^k}{\displaystyle r_0^{i+j+k}} 
\frac{\displaystyle \prod_{p=1}^{i/2} (2p-1) \prod_{q=1}^{j/2} (2q-1) \prod_{s=1}^{k/2} (2s-1)}{\displaystyle \prod_{u=1}^{(i+j+k)/2+2} (2u-1)} \\
& = \frac{3}{4\pi} \frac{\displaystyle a^i b^j c^k}{\displaystyle r_0^{i+j+k}} 
\dfrac{\Gamma\left(\dfrac{i+1}{2}\right)\Gamma\left(\dfrac{j+1}{2}\right)\Gamma\left(\dfrac{k+1}{2}\right)}{\Gamma\left(\dfrac{i+j+k+5}{2}\right)}
\end{split}
\label{eq_triaxial_Nijk}
\end{align}
if $i,j,k$ are even, and is zero otherwise.
This can be expanded to obtain the integral:
\begin{align}
\begin{split}
  & \int_{V} \dfrac{x^{i+i_\varrho} y^{j+j_\varrho} z^{k+k_\varrho}}{r_0^{i+i_\varrho+j+j_\varrho+k+k_\varrho}} \ \dfrac{{\text d}V}{r_0^3} \\
= & \dfrac{a^{i+i_\varrho+1} b^{j+j_\varrho+1} c^{k+k_\varrho+1}}{r_0^{i+i_\varrho+j+j_\varrho+k+k_\varrho+3}}
\dfrac{\Gamma\left(\dfrac{i+i_\varrho+1}{2}\right)\Gamma\left(\dfrac{j+j_\varrho+1}{2}\right)\Gamma\left(\dfrac{k+k_\varrho+1}{2}\right)}{\Gamma\left(\dfrac{i+i_\varrho+j+j_\varrho+k+k_\varrho+5}{2}\right)}
\end{split}
\end{align}
if $i+i_\varrho,j+j_\varrho,k+k_\varrho$ are even, and is zero otherwise.

\subsection{Spherical Harmonics}
\label{sec:SMrepSH}

A square-integrable scalar function $r(\theta,\phi)$
describing the surface of a planetary body
can be expanded using the spherical harmonics orthogonal basis
to obtain:
\begin{align}
r(\theta,\phi) & = \sum_{l=0}^{l_{\max}} \sum_{m=0}^{l} \left[ A_{lm} \cos(m\phi) + B_{lm} \sin(m\phi) \right] P_{lm}(\cos(\theta))
\label{eq:rtf}
\end{align}
where $r,\theta,\phi$ are the spherical coordinates radius, co-latitude, and longitude, respectively,
and $A_{lm},B_{lm}$ are the coefficients of the expansion.
The full spherical harmonics $Y_{lm}$ have already been converted in Eq.~\eqref{eq:rtf} using
$Y_{lm} = \bar P_{lm} \cos(m\phi)$ for $m\geq0$ and
$Y_{lm} = \bar P_{l|m|} \sin(|m|\phi)$ for $m<0$,
with the normalized associated Legendre functions 
$\bar P_{lm}(x) = P_{lm}(x) / N_{lm}$.
When the shape $r(\theta,\phi)$ is an observed quantity,
the coefficients of the spherical harmonics expansion can be obtained with the integrals 
\citep{1995geph.conf....1Y}:
\begin{align}
\label{eq:detA} A_{lm} & = \frac{1}{4\pi N_{lm}^2} \int r(\theta,\phi) P_{lm} (\cos(\theta)) \cos(m\phi) \ {\text d}\Omega \\
\label{eq:detB} B_{lm} & = \frac{1}{4\pi N_{lm}^2} \int r(\theta,\phi) P_{lm} (\cos(\theta)) \sin(m\phi) \ {\text d}\Omega
\end{align}
with $N_{lm}$ from Eq.~\eqref{eq:norm}.
This can be verified using:
\begin{align}
\int_{-1}^1 P_{lm}(x) P_{l'm}(x) \ \text{d}x & = \frac{2}{(2l+1)}\frac{(l+m)!}{(l-m)!} \delta_{ll'}
\end{align}
and 
\begin{align}
\int_0^{2\pi} \cos(m\phi) \cos(m'\phi) \ \text{d}\phi & = \frac{2\pi}{(2-\delta_{0m})} \delta_{mm'} \\
\int_0^{2\pi} \sin(m\phi) \sin(m'\phi) \ \text{d}\phi & = \frac{2\pi}{(2-\delta_{0m})} \delta_{mm'} 
\end{align}
also confirming indirectly the fact that we are using the so called ``geodesy'' or ``$4\pi$'' normalization \citep{Wieczorek_2007}:
\begin{align}
\int_\Omega Y_{lm}(\Omega) Y_{l'm'}(\Omega) \ {\text d}\Omega = 4\pi \delta_{ll'} \delta_{mm'} \ .
\end{align}

Additionally we note that the coefficients of the expansion are typically provided in their normalized
form
$\{\bar{A}_{lm},\bar{B}_{lm}\} = N_{lm} \{A_{lm},B_{lm}\}$.
This has the advantage of dealing with coefficients with comparable magnitude.

In the remainder of this section we will use powers of the radius function
$r(\theta,\phi)$ 
and the treatment is greatly simplified if we can first expand $r(\theta,\phi)$ as follows:
\begin{align}
\begin{split}
 & r(\theta,\phi) = \sum_{l=0}^{l_{\max}} \sum_{m=0}^{l} A_{lm} \cos(m\phi) P_{lm}(\cos(\theta)) + \sum_{l=0}^{l_{\max}} \sum_{m=0}^{l} B_{lm} \sin(m\phi) P_{lm}(\cos(\theta)) \\
= & \sum_{l=0}^{l_{\max}} \sum_{m=0}^{l} A_{lm} \sum_{\nu=0}^{\lfloor m/2 \rfloor} (-1)^\nu \binom{m}{2\nu} \cos^{m - 2 \nu}(\phi) \sin^{2\nu}(\phi) \\
  & 2^{-l} \sin^{m}(\theta) \sum_{u=0}^{\lfloor l/2 \rfloor} (-1)^u \binom{l}{u} \binom{2l-2u}{l} (l-m-2u+1)_{m} \cos^{l-m-2u}(\theta) \\
+ & \sum_{l=0}^{l_{\max}} \sum_{m=0}^{l} B_{lm} \sum_{\nu=0}^{\lfloor (m-1)/2 \rfloor} (-1)^\nu \binom{m}{2\nu+1} \cos^{m - 2 \nu - 1}(\phi) \sin^{2 \nu + 1}(\phi) \\
  & 2^{-l} \sin^{m}(\theta) \sum_{u=0}^{\lfloor l/2 \rfloor} (-1)^u \binom{l}{u} \binom{2l-2u}{l} (l-m-2u+1)_{m} \cos^{l-m-2u}(\theta) \\
= & \sum_{\tau=0}^1 \sum_{l=0}^{l_{\max}} \sum_{m=0}^{l} \sum_{u=0}^{\lfloor l/2 \rfloor} \sum_{\nu=0}^{\lfloor (m-\tau)/2 \rfloor} A_{lm}^{1-\tau} B_{lm}^{\tau} (-1)^{u+\nu} \binom{m}{2\nu+\tau} 2^{-l} \binom{l}{u} \binom{2l-2u}{l} \\
  & (l-m-2u+1)_{m} \cos^{m - 2 \nu - \tau}(\phi) \sin^{2\nu+\tau}(\phi) \cos^{l-m-2u}(\theta) \sin^{m}(\theta) \\
= & \sum_{\tau=0}^1 \sum_{l=0}^{l_{\max}} \sum_{m=0}^{l} \sum_{u=0}^{\lfloor l/2 \rfloor} \sum_{\nu=0}^{\lfloor (m-\tau)/2 \rfloor} A_{lm}^{1-\tau} B_{lm}^{\tau} Q_{\tau lmu\nu} \cos^{m - 2 \nu - \tau}(\phi) \sin^{2\nu+\tau}(\phi) \cos^{l-m-2u}(\theta) \sin^{m}(\theta)
\end{split}
\label{eq:rtf_expanded}
\end{align}
where 
the the sum over $\tau$ is introduced to have a single expansion for both $A_{lm}$ and $B_{lm}$ terms,
and we have have used:
\begin{align}
\cos(m \phi) & = \sum_{\nu=0}^{\lfloor m/2 \rfloor} 
(-1)^\nu \binom{m}{2\nu}
\cos^{m - 2 \nu}(\phi)
\sin^{2\nu}(\phi) 
\\ 
\sin(m \phi) & = \sum_{\nu=0}^{\lfloor (m-1)/2 \rfloor}   
(-1)^\nu \binom{m}{2\nu+1}
\cos^{m - 2 \nu - 1}(\phi)
\sin^{2 \nu + 1}(\phi)
\end{align}
and
\begin{align}
\label{eq:Plmdef}
P_{lm}(\cos(\theta)) = 2^{-l} \sin^{m}(\theta)
\sum_{u=0}^{\lfloor l/2 \rfloor} 
(-1)^u \binom{l}{u} \binom{2l-2u}{l} 
(l-m-2u+1)_{m} \cos^{l-m-2u}(\theta)
\end{align}
where the notation $\lfloor a \rfloor$ represents the floor of $a$,
and $(a)_{m}$ is the Pochhammer function of $a$.
Note that the term $(-1)^m$, used by some authors, is missing in Eq.~\eqref{eq:Plmdef}.
The coefficient $Q_{\tau lmu\nu}$ is defined as:
\begin{align}
Q_{\tau lmu\nu}=(-1)^{u+\nu} \binom{m}{2\nu+\tau} 2^{-l} \binom{l}{u} \binom{2l-2u}{l} (l-m-2u+1)_{m}
\end{align}

Now we use Eq.~\eqref{eq:rtf_expanded} to compute:
\begin{align}
\label{eq:pow_rn}
\begin{split}
  r^{N_r}(\theta,\phi)
= & \sum_{q_{00000}+\cdots+q_{\{\tau lmu \nu\}_{\max}}=N_r} \frac{N_r!}{q_{00000}! \cdots q_{\{\tau lmu \nu\}_{\max}}!} \\ 
  & \prod_{\tau_s l_s m_s u_s \nu_s=00000}^{\{\tau lmu \nu\}_{\max}} \left[ A_{l_s m_s}^{1-\tau_s} B_{l_s m_s}^{\tau_s} Q_{\tau_s l_s m_s u_s \nu_s} \right]^{q_{\tau_s l_s m_s u_s \nu_s}} \\ 
  & \prod_{\tau_s l_s m_s u_s \nu_s=00000}^{\{\tau lmu \nu\}_{\max}} \left[ \cos^{m_s - 2 \nu_s - \tau_s}(\phi) \sin^{2\nu_s+\tau_s}(\phi) \right]^{q_{\tau_s l_s m_s u_s \nu_s}} \\
  & \prod_{\tau_s l_s m_s u_s \nu_s=00000}^{\{\tau lmu \nu\}_{\max}} \left[ \cos^{l_s-m_s-2u_s}(\theta) \sin^{m_s}(\theta) \right]^{q_{\tau_s l_s m_s u_s \nu_s}}
\end{split}
\end{align}
where the product symbols are repeated three times to stress the fact that constant factors and variables depending on $\theta$ and $\phi$ are independent.
The expression is based on the general multinomial relation \citep{NIST_DLFM_book}:
\begin{equation}
\left[ \sum_{k=1}^{m} a_k \right]^n = \sum_{k_1+k_2+\cdots+k_m=n} \dfrac{n!}{k_1! k_2! \cdots k_m!} \prod_{\tau=1}^{m} a_\tau^{k_\tau}
\end{equation}
Working in spherical coordinates,
we can now compute the volume integral:
\begin{align}
\label{eq:rn_final}
\begin{split}
  & \int_{V} \dfrac{x^{i+i_\varrho} y^{j+j_\varrho} z^{k+k_\varrho}}{r_0^{i+i_\varrho+j+j_\varrho+k+k_\varrho}} \ \dfrac{{\text d}V}{r_0^3} \\
= & \int_{V} \dfrac{(r \sin(\theta) \cos(\phi))^{i+i_\varrho} (r \sin(\theta) \sin(\phi))^{j+j_\varrho} (r \cos(\theta))^{k+k_\varrho}}{r_0^{i+i_\varrho+j+j_\varrho+k+k_\varrho}} \dfrac{r^2}{r_0^2} \sin(\theta) \ \dfrac{{\text d}r}{r_0} {\text d}\theta {\text d}\phi \\
= & \frac{1}{N_r} \int_{\Omega} \dfrac{r^{N_r}}{r_0^{N_r}} \cos^{i+i_{\mathbf{w}}}(\phi) \sin^{j+j_{\mathbf{w}}}(\phi) \cos^{k+k_{\mathbf{w}}}(\theta) \sin^{i+i_{\mathbf{w}}+j+j_{\mathbf{w}}+1}(\theta) \ {\text d}\theta {\text d}\phi \\
= & \frac{1}{N_r} \sum_{q_{00000}+\cdots+q_{\{\tau lmu \nu\}_{\max}}=N_r} \frac{N_r!}{q_{00000}! \cdots q_{\{\tau lmu \nu\}_{\max}}!} \\
  & \prod_{\tau_s l_s m_s u_s \nu_s=00000}^{\{\tau lmu \nu\}_{\max}} \left[ \dfrac{A_{l_s m_s}^{1-\tau_s} B_{l_s m_s}^{\tau_s}}{r_0} Q_{\tau_s l_s m_s u_s \nu_s} \right]^{q_{\tau_s l_s m_s u_s \nu_s}} \\
  & \int_0^{2\pi} \prod_{\tau_s l_s m_s u_s \nu_s=00000}^{\{\tau lmu \nu\}_{\max}} \left[ \cos^{m_s - 2 \nu_s - \tau_s}(\phi) \sin^{2\nu_s+\tau_s}(\phi) \right]^{q_{\tau_s l_s m_s u_s \nu_s}} \cos^{i+i_\varrho}(\phi) \sin^{j+j_\varrho}(\phi) \ {\text d}\phi \\
  & \int_0^{\pi} \prod_{\tau_s l_s m_s u_s \nu_s=00000}^{\{\tau lmu \nu\}_{\max}} \left[ \cos^{l_s-m_s-2u_s}(\theta) \sin^{m_s}(\theta) \right]^{q_{\tau_s l_s m_s u_s \nu_s}} \cos^{k+k_\varrho}(\theta) \sin^{i+i_\varrho+j+j_\varrho+1}(\theta) \ {\text d}\theta 
\end{split}
\end{align}
where $N_r=i+i_\varrho+j+j_\varrho+k+k_\varrho+3$,
and the use of Eq.~\eqref{eq:pow_rn} allows us to separate of the two integrals in $\text{d}\phi$ and $\text{d}\theta$,
which are now in the form 
$\int_0^{k \pi} \cos^c(x) \sin^s(x) \ \text{d}x$ with solution:
\begin{equation}
\int_0^{k \pi}  \cos^c (x) \sin^s (x) \ \text{d}x =  
T_{csk}
\dfrac{\Gamma\left(\dfrac{c+1}{2}\right) \Gamma\left(\dfrac{s+1}{2}\right)}{\Gamma\left(\dfrac{c+s+2}{2}\right)}
\end{equation}
where $c,s,k$ are natural numbers including zero,
and the factor $T_{csk}$ is given by the following table:
\begin{center}
\begin{tabular}{ccc|c}
 $c$ &  $s$ &  $k$ & $T_{csk}$ \\
\hline
even & even &  any & $k$ \\
even &  odd &  odd &  1  \\
even &  odd & even &  0  \\
 odd &  any &  any &  0  \\
\end{tabular}
\end{center}
By using Eq.~\eqref{eq:rn_final} we can now compute the GMoI of a body with
arbitrary shape described using spherical harmonics, and arbitrary mass density distribution.
Note how the terms $i,j,k$ in Eq.~\eqref{eq:rn_final} appear explicitly only in the two
integrals, which makes it possible to parallelize the computation of the volume integral 
over all terms with constant $N_r$.

We note that a similar derivation was used in \cite{1994CeMDA..60..331B} in the context of obtaining the 
gravitational field of a homogeneous body,
while our method allows to model bodies with arbitrary mass density distribution.

\subsection{Latitude-Longitude Grid}
\label{sec:SMrepLatLon}

This is a sub-case of the more generic triangular mesh case described in \S\ref{sec:SMrepMesh},
as the points in the latitude-longitude grid can be grouped appropriately in sets of three
nearby points.

\subsection{Triangular Mesh}
\label{sec:SMrepMesh}

Polyhedral shapes have been used frequently to model planetary bodies with irregular shapes,
and to describe the interface between two layers or components
\citep{1976Geop...41.1353B,1989JGR....94.7555R,1997CeMDA..65..313W,2000P&SS...48..965S,2007Icar..192..150H}.
When the shape of a body is modeled using a triangular mesh,
the computation of the volume integral 
can be reduced to a sum over each simplex, 
where a simplex in three dimensional space is a tetrahedron with three vertices on the surface and one at the origin.
Approximate quadrature formulae exist, see i.e.~\cite{1978SJNA...15..282G}
which provide distinct quadrature rules for polynomials up to a given degree.
More recently, \cite{SimplexIntegration} have derived the general exact formulas,
where the function is evaluated only at the vertices of the simplices.
We refer the reader to these two articles for the implementation details.

\begin{table*}
\begin{center}
\begin{tabular}{cc|l|l}
$l$ & $m$ & $\bar{C}_{lm}$ & $\bar{S}_{lm}$ \\ 
\hline
0 & 0 & ${\cal N}_{000}$ & --- \\
1 & 0 & ${\cal N}_{001}/\sqrt{3}$ & --- \\
1 & 1 & ${\cal N}_{100}/\sqrt{3}$ & ${\cal N}_{010}/\sqrt{3}$ \\
2 & 0 & $(2 {\cal N}_{002} - {\cal N}_{020} - {\cal N}_{200})/(2\sqrt{5})$ & --- \\
2 & 1 & $\sqrt{3/5} {\cal N}_{101}$ & $\sqrt{3/5} {\cal N}_{011}$ \\
2 & 2 & $\sqrt{3/20} ({\cal N}_{200} - {\cal N}_{020})$ & $\sqrt{3/5} {\cal N}_{110}$ \\
3 & 0 & $(2 {\cal N}_{003} - 3 {\cal N}_{021} - 3 {\cal N}_{201})/(2\sqrt{7})$ & --- \\
3 & 1 & $\sqrt{3/56} (4 {\cal N}_{102} - {\cal N}_{120} - {\cal N}_{300})$ & $\sqrt{3/56} (4 {\cal N}_{012} - {\cal N}_{030} - {\cal N}_{210})$\\
3 & 2 & $\sqrt{15/28} ({\cal N}_{201} - {\cal N}_{021})$ & $\sqrt{15/7} {\cal N}_{111}$ \\
3 & 3 & $\sqrt{5/56} ({\cal N}_{300} - 3 {\cal N}_{120})$ & $\sqrt{5/56} (3 {\cal N}_{210} - {\cal N}_{030})$ \\
4 & 0 & $(8 {\cal N}_{004} - 24 {\cal N}_{022} + 3 {\cal N}_{040} - 24 {\cal N}_{202} + 6 {\cal N}_{220} + 3 {\cal N}_{400})/24$ & --- \\
4 & 1 & $\sqrt{5/72} (4 {\cal N}_{103} - 3 {\cal N}_{121} -3 {\cal N}_{301})$ & $\sqrt{5/72} (4 {\cal N}_{013} - 3 {\cal N}_{031} - 3 {\cal N}_{211})$ \\
4 & 2 & $\sqrt{5}/12 ({\cal N}_{040} - {\cal N}_{400} + 6 {\cal N}_{202} - 6 {\cal N}_{022})$ & $\sqrt{5}/6 (6 {\cal N}_{112} - {\cal N}_{130} - {\cal N}_{310})$ \\
4 & 3 & $\sqrt{35/72} ({\cal N}_{301} - 3 {\cal N}_{121})$ & $\sqrt{35/72} (3 {\cal N}_{211} - {\cal N}_{031})$ \\
4 & 4 & $\sqrt{35}/24 ({\cal N}_{040} - 6 {\cal N}_{220} + {\cal N}_{400})$ & $\sqrt{35}/6 ({\cal N}_{310} - {\cal N}_{130})$ \\
\end{tabular}
\end{center}
\caption{
Coefficients from Eq.~\eqref{eq:CS} expressing the normalized coefficients of the spherical harmonics expansion of the gravitational 
potential $\bar{C}_{lm}$ and $\bar{S}_{lm}$ in therms of the generalized moments of inertia coefficients ${\cal N}_{ijk}$.
These expressions hold in any reference system, barycentric or not.
}
\label{tab:conv}
\end{table*}

\begin{table}
\begin{center}
\begin{tabular}{cc|r}
$l$ & $m$ & $\bar{A}_{lm}$ \\ 
\hline
0 & 0 &  57.0  \\
1 & 1 &   2.5  \\
2 & 0 & $-6.0$ \\
2 & 2 &   5.0  \\
3 & 1 & $-1.5$ \\
3 & 3 &   2.0  \\
4 & 2 & $-1.0$ \\
4 & 4 &   2.0  \\
5 & 3 & $-0.5$ \\
\end{tabular}
\end{center}
\caption{
Spherical harmonics expansion (see Eq.~\eqref{eq:rtf}) of the shape of the sample body (see \S\ref{sec:sample}).
All coefficients are normalized and are in km.
The coefficients not listed, including all the $\bar{B}_{lm}$ coefficients, are zero.
The total volume is $8.364117 \times 10^5$~km$^3$.}
\label{tab:SH_body_1}
\end{table}

\begin{table}
\begin{center}
\begin{tabular}{ccc|c}
$i$ & $j$ & $k$ & $\Phi_{ijk}$ \\ 
\hline
0 & 0 & 0 & 0.836411678 \\
1 & 0 & 0 & 0.068883083 \\
0 & 0 & 2 & 0.036327037 \\
0 & 2 & 0 & 0.048925999 \\
2 & 0 & 0 & 0.113678894 \\
1 & 0 & 2 & 0.002378213 \\
1 & 2 & 0 & 0.002275354 \\
3 & 0 & 0 & 0.032958810 \\
0 & 0 & 4 & 0.003254146 \\
0 & 2 & 2 & 0.001596560 \\
0 & 4 & 0 & 0.006328626 \\
2 & 0 & 2 & 0.003787446 \\
2 & 2 & 0 & 0.004178518 \\
4 & 0 & 0 & 0.036373279 \\
\end{tabular}
\end{center}
\caption{
Volume integrals as defined in Eq.~\eqref{eq:integral_def} of the body with shape 
from spherical harmonics coefficients in Table~\ref{tab:SH_body_1}.
These integrals are computed using Eq.~\eqref{eq:rn_final}
in the same body-fixed reference frame where the coefficients in Table~\ref{tab:SH_body_1} are defined.
Only the non-zero integrals up to degree 4 are included in this table.
The normalization radius has been arbitrarily chosen to be $r_0=100$~km.
}
\label{tab:shape_integral_body_1}
\end{table}

\begin{table}
\begin{center}
\begin{tabular}{rr|r|r|r|r}
    &     &     body-fixed &    barycentric &     body-fixed &    barycentric \\
    &     &        uniform &        uniform &      composite &      composite \\
$l$ & $m$ & $\bar{C}_{lm}$ & $\bar{C}_{lm}$ & $\bar{C}_{lm}$ & $\bar{C}_{lm}$ \\ 
\hline
0 & 0 &   1.000000  &   1.000000  &   1.000000  &   1.000000  \\
1 & 1 &   0.047548  &   0.000000  &   0.039545  &   0.000000  \\
2 & 0 & $-0.024048$ & $-0.022531$ & $-0.022405$ & $-0.021356$ \\
2 & 2 &   0.029984  &   0.027357  &   0.027566  &   0.025749  \\
3 & 1 & $-0.007118$ & $-0.001801$ & $-0.006359$ & $-0.002202$ \\
3 & 3 &   0.009336  &   0.003954  &   0.008290  &   0.004112  \\
4 & 0 &   0.002490  &   0.001703  &   0.002240  &   0.001609  \\
4 & 2 & $-0.003765$ & $-0.002545$ & $-0.003365$ & $-0.002396$ \\
4 & 4 &   0.005196  &   0.003402  &   0.004617  &   0.003221  \\
\end{tabular}
\end{center}
\caption{
Normalized spherical harmonics coefficients of the gravitational potential expansion
for the sample body density models described in \S\ref{sec:sample},
expanded about the origin (body-fixed) or
about the center of mass, which is 
at $(x,y,z) = (8.235548,0,0)$~km for the uniform case,
and at $(x,y,z) = (6.849403,0,0)$~km for the composite case.
Only the non-zero coefficients up to degree 4 are listed,
and all the $\bar{S}_{lm}$ coefficients are zero in this particular example.
}
\label{tab:gravity_body_1}
\end{table}

\end{document}